\documentclass[sigconf]{acmart}
\newcommand{\revised}[1]
{\textcolor{black}{#1}}
\usepackage{hyperref}
\usepackage{multirow}
\usepackage{hhline}
\usepackage{graphicx}
\usepackage{array}
\hypersetup{
    colorlinks=true,
    linkcolor=blue,
    filecolor=magenta,      
    urlcolor=cyan,
}
\AtBeginDocument{%
  }

\usepackage{amsmath}
\usepackage{booktabs}
\usepackage{multirow}
\usepackage{amsmath,amsfonts}
\usepackage{algorithmic}
\usepackage{algorithm}
\usepackage{array}
\usepackage[caption=false,font=normalsize,labelfont=sf,textfont=sf]{subfig}
\usepackage{textcomp}
\usepackage{stfloats}
\usepackage{url}
\usepackage{verbatim}
\usepackage{graphicx}
\usepackage{hhline}
\usepackage{makecell}
\usepackage{enumitem}

\newcolumntype{C}[1]{>{\centering\arraybackslash}m{#1}}

\setcopyright{acmlicensed}
\copyrightyear{2026}
\acmYear{2026}
\acmDOI{XXXXXXX.XXXXXXX}
\acmConference[CHI '26]{CHI}{April 13–17,
  2026}{Barcelona, Spain}
\acmBooktitle{CHI (CHI '26), April 13–17, 2026, Barcelona, Spain}
\acmISBN{978-1-4503-XXXX-X/2018/06}

\begin{document}

\title{Designing a Generative AI-Assisted Music Psychotherapy Tool for Deaf and Hard-of-Hearing Individuals}

\author{Youjin Choi}
\orcid{0000-0003-2788-7871}
\authornote{First author.}
\affiliation{%
  \institution{Department of AI Convergence, Gwangju Institute of Science and Technology}
  \city{Gwangju}
  \state{Buk-gu}
  \country{Korea, Republic of,}
}
\email{chldbwls304@gm.gist.ac.kr}

\author{JaeYoung Moon}
\orcid{0000-0002-1852-2769}
\affiliation{%
  \institution{Department of AI Convergence, Gwangju Institute of Science and Technology}
  \city{Gwangju}
  \state{Buk-gu}
  \country{Korea, Republic of,}
}
\email{super_moon@gm.gist.ac.kr}

\author{JinYoung Yoo}
\orcid{0009-0002-0751-4349}
\affiliation{%
  \institution{Department of AI Convergence, Gwangju Institute of Science and Technology}
  \city{Gwangju}
  \state{Buk-gu}
  \country{Korea, Republic of,}
}
\email{9inceroo@gm.gist.ac.kr}

\author{Jennifer G Kim}
\orcid{0000-0003-3253-3963}
\affiliation{%
  \institution{School of Interactive Computing, Georgia Institute of Technology}
  \city{Atlanta}
  \state{Georgia}
  \country{United States}
}
\email{jennifer.kim@cc.gatech.edu}

\author{Jin-Hyuk Hong}
\orcid{0000-0002-8838-5667}
\authornote{Corresponding author.}
\affiliation{%
  \institution{Department of AI Convergence, Gwangju Institute of Science and Technology}
  \city{Gwangju}
  \state{Buk-gu}
  \country{Korea, Republic of,}
}
\email{jh7.hong@gist.ac.kr}

\renewcommand{\shortauthors}{Choi et al.}

\begin{abstract}
Songwriting has long served as a powerful medium for expressing unconscious emotions and fostering self-awareness in psychotherapy. Due to the auditory-centric nature of traditional approaches, Deaf and Hard-of-Hearing (DHH) individuals have often been excluded from music’s therapeutic benefits. In response, this study presents a music psychotherapy tool co-designed with therapists, integrating conversational agents (CAs) and music generative AI as symbolic and therapeutic media. Through a usage study with 23 DHH individuals, we found that collaborative songwriting with the CA enabled them to experience emotional release, reinterpretation, and deeper self-understanding. In particular, the CA’s strategies—supportive empathy, example response options, and visual-based metaphors—were found to facilitate musical dialogue effectively for DHH individuals. These findings contribute to inclusive AI design by showing the potential of human–AI collaboration to bridge therapeutic and artistic practices.
\end{abstract}



\maketitle

\section{Introduction}
\label{sec: introduction}
Music has long been employed as a powerful medium for emotional regulation and self-awareness in psychotherapy \cite{Schäfer2013, Stewart2019, Cai2023, Jin2024}. From a psychodynamic view, it serves as a catalyst connecting individuals to their internal world, evoking personal memories and experiences, and aiding in recognizing and restructuring negative emotions \cite{Schäfer2013, wilkins2014, Welch2020, Donald2015}. Building on these characteristics, music therapy incorporates structured musical activities, such as listening \cite{ward2016}, improvisation \cite{Roberts2013}, or songwriting \cite{Baker01022011}, as therapeutic interventions to facilitate emotional exploration and healing. Among various interventions, songwriting can help uncover unconscious emotions or unresolved issues, offering a meaningful channel for self-awareness and emotional expression. These potentials have made songwriting a widely adopted form of emotional expression in clinical practice, including those with substance use disorders, adolescents, and cancer patients \cite{Silverman2012, Dvorak2016, jung2015}.

Despite its therapeutic promise, songwriting remains largely inaccessible to DHH individuals. Music can serve as a symbolic and meaningful form of emotional care for DHH individuals facing social isolation and psychological distress due to hearing loss \cite{Youjin2025IJHCS, Youjin2025CHI}; however, they have often been excluded from songwriting-based emotional care \cite{Darrow2006, ward2016}. Existing research on music therapy for DHH populations\revised{, particularly cochlear implant (CI) users, has largely focused on music perception training aimed at enhancing auditory function} \cite{Shukor2021, Mark2021, Jiam2019}. While this may support residual hearing use, it does not enable DHH individuals to access music’s emotional benefits, such as expression, reflection, and healing. This gap stems not only from auditory barriers but also from a lack of inclusive therapeutic and technological approaches that accommodate the sensory and emotional needs of DHH individuals. As a result, traditional auditory-centered music psychotherapy has not sufficiently enabled DHH individuals to explore music emotionally, highlighting the need for new technological interventions and a paradigm shift.

While the auditory nature of music has posed challenges for DHH individuals in therapeutic contexts, advances in multimodal interfaces have opened alternative pathways for musical engagement and expression \cite{ward2016}. Prior research increasingly shows that DHH individuals do not merely appreciate music but engage with it through multisensory modes, including visual, tactile, and narrative forms \cite{Youjin2025IJHCS, Youjin2025CHI, lin2022, derberg2016}. Recently, prompt-based music generative AI (GenAI), which enables music composition from natural-language prompts, has reduced barriers for DHH users and created new opportunities for self-expression \cite{Youjin2025CHI}. To translate these opportunities into therapeutic benefits, however, interactive guidance that ensures both sensory and emotional accessibility, such as adaptive prompts, multimodal feedback, and step-by-step scaffolding, is essential \cite{Sun2024}. Despite this need, intervention studies that support music therapy for DHH individuals remain scarce \cite{ward2016}.

Against this backdrop, our study \revised{investigates} how DHH individuals can engage in music co-creation with GenAI technologies, including music GenAI, an LLM-based CA, and music visualization, as a novel form of psychotherapy. \revised{We recruited Korean CI users who identify as Deaf\footnote{``Deaf'' means that they identify as part of the Deaf community \cite{Padden1988}}, live within a hearing society, regularly use text-based digital communication, and hold strong interests in music and music psychotherapy. These individuals represent probable early adopters of such a system in real-world therapeutic contexts.} We address the following research questions:
\begin{itemize}
    \item RQ1. What is the current state of therapeutic practices in music psychotherapy for DHH individuals?
    \item RQ2. \revised{What design principles should a GenAI-assisted songwriting tool follow to effectively support music psychotherapy for DHH individuals?}
    \item RQ3. \revised{How does collaborative songwriting using a GenAI-assisted tool affect the emotional and therapeutic outcomes of DHH individuals?}
\end{itemize}

To explore these questions, we first conducted in-depth interviews with six music therapists to identify key challenges and needs in music therapy for DHH clients (RQ1). Drawing on these insights, we held a co-design workshop with four music therapists, during which we identified key challenges, such as emotional withdrawal and lack of songwriting confidence among DHH individuals, \revised{and derived design principles for a GenAI-assisted songwriting tool that incorporates music therapy techniques and therapeutic conversation strategies to address these issues (RQ2).} We then ran a usage study with \revised{23 CI users who identify as Deaf} to examine how co-creative songwriting with the CA emotional expression and therapeutic engagement (RQ3). 
The results suggest that the CA’s use of supportive empathy, example response options, and visual-based metaphors and analogies enhanced participants’ capacity to express and share their personal stories through music. Notably, the tool's visual-based questions, which guided users to visualize music as images or scenes, played a critical role in enhancing musical expression. Beyond enabling songwriting, the tool showed therapeutic potential for \revised{this subgroup of DHH users} in fostering emotional awareness, expressive articulation, and psychological reframing. 

The contributions of this study are as follows:
\begin{itemize}
    \item We design and develop a GenAI-assisted music psychotherapy tool tailored to the characteristics of DHH individuals through a co-design workshop with professional therapists.
    \item Through a usage study, we demonstrate the potential of our GenAI-assisted tool to support therapists and enhance emotional care in health and well-being contexts. 
    \item Our findings lead to the design of emotionally supportive GenAI systems by demonstrating how such tools can facilitate meaningful collaboration between DHH users and CA, emphasizing strategies that promote self-disclosure and inclusive co-creation.
\end{itemize}


\section{Background}
\label{sec: background}
\subsection{Benefits of Music Psychotherapy}
Psychology and neuroscience research consistently shows that music positively influences the brain and cognitive functioning, contributing to improved mental health and well-being \cite{Schäfer2013, wilkins2014, Welch2020, Donald2015}. Music activates neural networks, evokes memories, and supports emotional recognition and regulation \cite{Schäfer2013, Stewart2019, Cai2023, Jin2024}. Along with these emotional effects, music psychotherapy has incorporated various musical activities, such as listening, composing, and singing, that activate memories and emotions and support emotional healing \cite{Silverman2012, mcferran2006, Clare2009, Roberts2013}. While various music-based interventions show therapeutic benefits, songwriting specifically offers unique advantages that other activities cannot provide \cite{jung2015, Roberts2013}. Unlike passive forms of engagement with music, the act of creating music engages individuals as active agents in their own therapeutic process, fostering a sense of ownership and empowerment \cite{jung2015, Roberts2013, Felicity2011}. Songwriting allows for personalized emotional expression through artistic choices that transcend verbal limitations, making it particularly valuable for those who struggle with direct emotional articulation \cite{Donna2022, Silaji2024, Matthew2020}. Therefore, it has been implemented across diverse clinical contexts with clients who face challenges in self-expression, including those with substance use disorders, adolescents, and cancer patients \cite{Silverman2012, Callaghan2009, Dalton2006}.

\revised{Despite its therapeutic potential, music psychotherapy has largely excluded DHH individuals because of auditory constraints \cite{ward2016}. When DHH people do receive music therapy, most often as CI or hearing-aid users, the interventions typically focus on improving auditory access and perceptual skills (\textit{e.g.,} sound detection, discrimination, and speech perception) rather than supporting emotional well-being \cite{Shukor2021, Mark2021, Jiam2019}. Consequently, interventions explicitly oriented toward emotional well-being have remained limited in both scope and application. DHH individuals, however, often experience emotional distress related to auditory deprivation and social exclusion, and songwriting can help address these emotional needs \cite{moores2001, meadow1980}. Building on this clinical context, our study reframes music psychotherapy as a multimodal space for emotional expression rather than as a form of auditory rehabilitation.} While conventional auditory-based interventions may present accessibility barriers, the creative process of songwriting provides a more inclusive pathway to emotional expression. By acknowledging their unique sensory modalities, including visual, tactile, and other non-auditory channels, a multimodal approach to music psychotherapy can enable DHH individuals to more fully benefit from music’s emotional potential \cite{Youjin2025IJHCS, Youjin2025CHI}.

\subsection{Music and DHH Individuals}
Music is not solely an auditory experience; it is a multisensory art form encompassing narrative, visual, and tactile elements \cite{churchill2016, whittaker2008, cheng2016}. This broad definition challenges the conventional assumption that music belongs exclusively to non-DHH individuals, instead highlighting the ability of DHH individuals to engage with music in unique and meaningful ways. Churchill \cite{churchill2016} addressed DHH musical experiences from a cultural perspective, proposing a shift from sound-based definitions to those emphasizing narrative and multimodal embodiment. In practice, many DHH artists engage with music through full-body sensory experiences, interpreting rhythm and emotion via visual and tactile feedback, and contribute as sound designers, vocalists, dancers, and more \cite{Sylvain2021, Ohshiro2024}. DHH individuals often experience music through multimodal channels such as song signing \cite{Youjin2025IJHCS}, captioned music videos \cite{Vy2008, Lee2002}, and haptic devices \cite{La2014, Karam2010}. These approaches have informed the development of music-sensory substitution systems, which enhance or replace auditory components with alternative sensory cues. Such systems have significantly improved DHH users’ access to listening, performance, and songwriting \cite{lin2022, derberg2016, Petry2018, Youjin2024}. In parallel, technology has increasingly supported music creation and expression among DHH individuals. For instance, Tony \cite{lin2022} developed a system that conveys rhythm and beat through visual and haptic cues, while Søderberg \textit{et al.} \cite{derberg2016} introduced a visual interface for collaborative rhythm generation between DHH and hearing users. Choi \textit{et al.} \cite{Youjin2025IJHCS} designed a multimodal synthesizer that enables collaborative song signing. Their subsequent study proposed a prompt-based music GenAI tool, allowing DHH users to independently create narrative-driven music without relying on hearing collaborators \cite{Youjin2025CHI}.

These studies offer empirical evidence that, with appropriate technological support, DHH individuals can actively engage in music appreciation and creation. Notably, they affirm that music transcends sensory experience, functioning as a powerful medium for emotional expression and healing. These findings suggest that music psychotherapy, supported by AI-based tools that offer both accessibility and multi-modal functionality, holds significant therapeutic potential for individuals with hearing impairments. 

\subsection{Generative AI Technology in Psychotherapy}
With the rise in emotional challenges such as anxiety, depression, and social isolation, the Human Computer Interaction (HCI) community has developed various technologies to support mental well-being \cite{David2018, Kocielnik2018, choi2025private}. Recent studies have utilized GenAI, particularly LLM-based CAs and creative GenAI systems, to promote emotional recognition, expression, and resilience.

\textbf{LLM-based CA}. LLM-based CAs have gained attention as tools for journaling, emotional expression, and self-reflection \cite{Taewan2024, Inhwa2025, choi2025private}. These systems ease the psychological burden of disclosure and enable ongoing emotional engagement across time and space. They can also implement diverse therapeutic strategies, including expressive writing, reinforcement, and empathetic dialogue \cite{Seo2024, Taewan2024}. CAs are typically used in two ways in psychotherapy. First, they serve as tools for therapists. For example, \textit{MindfulDiary} \cite{Taewan2024} analyzes emotional logs to provide insights that improve therapy and \textit{Psy-LLM} \cite{lai2023} generates therapeutic dialogues using Q\&A data and clinical literature to assist mental health professionals. Second, CAs directly support clients as self-expression tools. For instance, Song \textit{et al.} \cite{Inhwa2025} include features like keyword prompts, theme generation, and AI-assisted summaries to help users articulate inner thoughts and maintain journaling. Liu \textit{et al.} \cite{Liu2024} identifies emotional states and challenges, providing empathetic, personalized responses that foster self-awareness and emotional regulation. These tools are especially useful when users face social isolation or lack access to traditional mental health services.

\textbf{Creative GenAI}. In parallel, the use of creative GenAI has recently expanded in art-based psychotherapy, including visual art, music, and creative writing \cite{Jin2024, Sun2024, Eunseo2022}. These tools reduce psychological barriers related to skill, creative anxiety, or communication difficulties. For instance, \cite{Xuejun2024} developed \textit{DeepThink} to support visual art creation without requiring professional expertise. This system assists with color selection, pattern generation, and image transformation during therapy. In music therapy, creative GenAI applications are also emerging. Sun \textit{et al.} \cite{Sun2024} examined music therapists’ perspectives on using Music GenAI for emotion regulation or personalized music recommendations. Jin \textit{et al.} \cite{Jin2024} explored older adults’ views on using creative GenAI in music reminiscence therapy and outlined design considerations for clinical settings. These studies suggest that music GenAI has therapeutic promise for DHH individuals, particularly when focusing on sensory substitution and narrative-driven composition. When effectively integrated, such tools have the potential to translate emotional states into personalized musical narratives that are accessible regardless of auditory ability.

Building on this foundation, our study explores the role of CAs and music GenAI not only in supporting therapists’ treatment processes for DHH clients but also in designing systems that provide entry-level music therapy tasks accessible to them.

\section{Co-designing an AI-assistive Music Psychotherapy Tool (RQ1 and RQ2)}
\label{sec: co-creative design workshop}
As GenAI becomes more prevalent in mental health due to its accessibility and user-centered design, effective clinical use requires customization based on therapeutic domains, user needs, and goals. Co-design with therapists is essential to ensure both usability and clinical relevance \cite{Taewan2024, Inhwa2025}. At the same time, because LLM-based CAs may generate inconsistent or harmful content that compromises therapeutic safety, domain-informed dialogue flow designs are critical for responsible and effective integration into psychotherapy \cite{Dasom2024, Kong2025}.

To address these challenges, we adopted a co-design approach with professional music therapists to explore the use of music from a psychotherapy perspective. DHH users were not included in this initial phase, as the primary aim was to translate established therapeutic techniques into a structured framework for the tool. We first conducted individual interviews with six music therapists to examine the current state of music psychotherapy for DHH clients and to identify needs for supportive tools (section~\ref{individual interview}). Subsequently, we collaborated with four therapists in a creative design workshop to co-develop the therapeutic process and define the core functionalities of a GenAI-assisted music psychotherapy tool (section~\ref{co-design workshop}).

\subsection{Individual Interviews}
\label{individual interview}
\subsubsection{Interview Design}\hfill

\textbf{Participants.} \revised{To address RQ1 on the current state of music psychotherapy practices for DHH individuals, we} first conducted individual interviews with six licensed music therapists who had experience working with DHH clients. Eligible participants were required to have (1) over five years of clinical experience and (2) a relevant professional qualification, such as licensure in counseling or a doctorate in clinical psychology. Therapists without prior experience working with DHH clients were excluded because music therapy for DHH individuals demands specialized knowledge and protocols tailored to their unique sensory and emotional characteristics. Therefore, recruiting qualified participants was challenging \cite{ward2016, Palmer2022}; therefore, six participants were recruited for this study. All therapists were recruited via professional networks and referrals from therapeutic centers. All participants identified as female. Their age (M=36.5, SD=3.78) and years of experience (M=7.83, SD=2.78) are presented in (Table~\ref{tab: population}). Each therapist had worked with a broad range of DHH clients, from children to older adults. While their primary field was music perception, some also incorporated music psychotherapy in clinical contexts. All participants were familiar with GenAI tools such as ChatGPT \footnote{https://chatgpt.com/} and had hands-on experience using LLM-based CAs or music GenAI systems like SUNO\footnote{https://suno.com/}. Before the interviews, participants completed a demographic survey and provided written consent. The study was approved by the Institutional Review Board (IRB) at [anonymized university], Korea.

\begin{table*}[t]
\begin{center}
\caption{Demographic information of the individual interview and design workshop}
  \label{tab: population}
  \begin{tabular}{c|c|c|c|c|c|c|c|c}
    \toprule
    \multirow{2}{*}{ID} & \multirow{2}{*}{Gender} & \multirow{2}{*}{Age} & \multirow{2}{*}{Job title} & \multirow{2}{*}{Specific field} & Years of & \multicolumn{2}{c|}{Participation} & \multirow{2}{*}{\revised{Knowledge of GenAI}}\\
    \hhline{~~~~~~--|~}
    &&&&& experience & Interview & DW & \\
    \midrule
    T1 & Female & 40 & Clinical music therapist & Percept (Psycho) & 11 & O & O & Experienced\\
    T2 & Female & 32 & Licensed music therapist & Percept (Psycho) & 5 & O & O & Experienced\\
    T3 & Female & 35 & Licensed music therapist & Percept (Psycho) & 6  & O & O & Experienced\\
    T4 & Female & 36 & Licensed music therapist & Percept (Psycho) & 10 & O & O & Knowledge\\
    T5 & Female & 42 & Clinical music therapist & Percept & 10 & O &  - & Experienced\\
    T6 & Female & 34 & Licensed music therapist & Percept & 5 & O & - & Knowledge\\
    \bottomrule
    \multicolumn{9}{>{\raggedright\arraybackslash}p{\dimexpr\linewidth-2\tabcolsep\relax}}{\small Job title: someone who holds a music therapy-related license (Licensed music therapist) and either has a doctoral degree as a clinical music therapist (Clinical music therapy); Specific field: primarily music perception (Percept), with additional experiences in music psychotherapy (Psycho); Participation: interview and design workshop (DW); \revised{GenAI: prior usage (Experienced) and awareness only (Knowledge)}.}
\end{tabular}
 \vspace{-10pt}
\end{center}
\end{table*}

\textbf{Process.} We conducted 60-minute semi-structured interviews focused on two main topics: (1) the current state of music psychotherapy for DHH clients, and (2) the practical needs and challenges in this domain (see Appendix~\ref{adx: interview details} for more details). With consent, all sessions were recorded and transcribed. Two researchers independently coded the transcripts by categorizing data into three themes: therapeutic context, challenges, and support needs using thematic coding \cite{Virginia2006}. After two rounds of discussion, the coding was refined, and the following two key findings were identified.

\subsubsection{Findings. Current Practices and Needs in Music Psychotherapy for DHH individuals \revised{(RQ1)}} \hfill

\textbf{Therapeutic priorities shaped by constraints of time and cost.} 
Therapists reported that most DHH clients prioritize music perception training aimed at improving auditory function over long-term emotional healing, largely due to constraints in time, cost, and accessibility. These limitations were especially pronounced among adult clients, who often favored short-term, results-oriented auditory training. Many believed that improving their auditory perception through music-based training would naturally resolve their social and emotional difficulties. 

\begin{quote}
\textit{``From the client’s perspective, recovering auditory function is observed as the top priority. Emotional therapy often gets pushed aside, especially for adults who are constrained by time and cost due to work or life responsibilities.''} (T1)
\end{quote}
\begin{quote}
\textit{``Adults tend to be reluctant to invest in emotional therapy. They want to feel their time and money were well spent—usually through functional outcomes.''} (T4)
\end{quote}

However, therapists noted that while perception-based approaches had functional benefits, they often failed to address underlying emotional needs. Hearing loss is often described as a \textit{“hidden disability,”} contributing to frequent misunderstandings and feelings of exclusion among DHH individuals. Therapists unanimously noted that their DHH clients frequently struggled with chronic emotional issues, including low self-esteem and depression. Even when their auditory perception improved through training, therapists consistently observed symptoms of depression, anxiety, and stress rooted in social isolation or unresolved trauma. Even with improved auditory skills, many clients continued to experience emotional distress and lacked confidence in social settings.
\begin{quote}
\textit{``Even after regaining some hearing ability, many clients carry self-defeating beliefs like `I’m still not good enough.' It’s like treating a car accident injury without addressing the trauma.''} (T6)
\end{quote}

\textbf{Symbolic significance of music to DHH individuals and the lack of tailored psychotherapy tools.} To address this, four therapists integrated songwriting into treatment alongside perception training, using music and lyrics as tools for emotional expression and self-reflection. They shared examples where songwriting enables clients to express internal struggles and rebuild confidence. However, therapists noted challenges in consistently applying songwriting, especially with adult clients facing a shorter treatment window. While therapists sought remote or digital solutions to extend emotional support outside of sessions, they emphasized the lack of therapy systems that account for the unique sensory and communicative characteristics of DHH individuals. In particular, emotional engagement through music listening remains constrained for many DHH clients, underscoring the need for more accessible and multimodal systems. All therapists emphasized the need for accessible and flexible systems to support emotional healing through music, especially for DHH individuals facing practical and emotional barriers to in-person care. 
\begin{quote}
\textit{``With a postlingual hearing loss client, we used song discussion for about three months. Eventually, they shared that colleagues noticed he looked more cheerful and spoke more confidently at work.''} (T2)
\end{quote}
\begin{quote}
\textit{``There are many clients who can’t afford therapy or avoid social interactions, so we need remote-access tools. But emotional therapy systems are nearly non-existent compared to music perception tools.''} (T3)
\end{quote}

\subsection{Co-design Workshop}
\label{co-design workshop}
Insights from interviews revealed key challenges in DHH music psychotherapy: cost constraints and insufficient systemic support. These barriers often prevent consistent access to psychotherapeutic services, especially without the presence of a therapist. To address this gap, we explored the potential of a GenAI-assisted music psychotherapy support tool that could offer greater flexibility and accessibility. We conducted a co-design workshop with four of them to explore a GenAI-based intervention. \cite{Norman1986}.

\subsubsection{Participants and Process}
\hfill

\textbf{Participants.} The co-design workshop \revised{included} practicing therapists at an early stage of the design process, ensuring that the GenAI tool reflects therapeutic perspectives while addressing the needs and values of DHH users.  Four music therapists (T1-T4) who had participated in the prior interviews were recruited; they had experience treating DHH clients with anxiety, stress, or depression (Table~\ref{tab: population}). \revised{The lead therapist (T1) was selected from this cohort based on more than ten years of clinical work with DHH clients and prior hands-on experience with LLM-based CAs and music-generation tools, for example, ChatGPT and SUNO. Before data collection, we held a 30-minute preparation session with the lead therapist to review the workshop goals, ethical considerations, and risk-management procedures. Because therapists differed in familiarity with GenAI, we ran a brief hands-on exercise using ChatGPT and SUNO for all participants and collected short written reflections on perceived affordances prior to the main ideation activities.} 

\begin{figure*}[ht]
  \centering
  \includegraphics[width=1\textwidth]{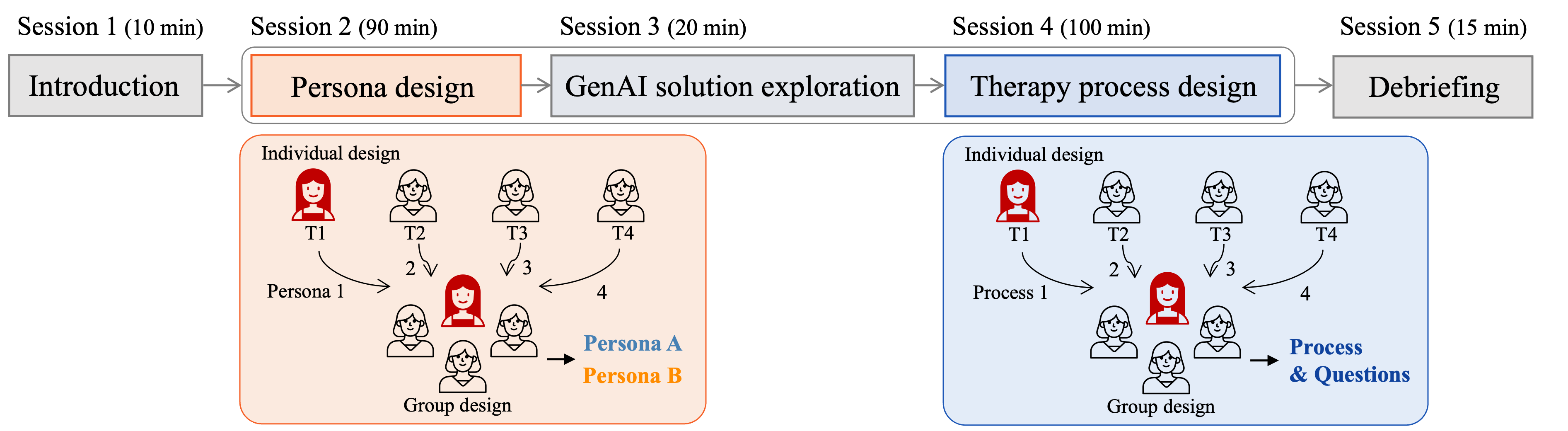}
  \caption{Design workshop process. Sessions comprise persona design, a discussion about GenAI-based technology, and process design. In the design session, the final design emerged from group discussion, led by the main facilitator (T1, red), after individual design work.}
  \label{fig: design workshop process}
\end{figure*}

\textbf{Process.} Following user-centered design principles \cite{Norman1986}, we conducted a 4-hour remote workshop comprising five sessions (Figure~\ref{fig: design workshop process}): (1) introduction, (2) persona design, (3) GenAI solution exploration, (4) therapy process design, and (5) debriefing. To tailor the tool to the needs of DHH clients, we employed a persona design approach that allowed therapists to externalize their tacit knowledge and articulate user needs more concretely \cite{Norman1986, Weitz2024}. Participants first created representative personas based on their clinical experiences, grounding the design in real-world therapeutic practice (persona design). They then identified expected challenges in applying GenAI-based interventions to these personas and explored potential solutions (GenAI solution exploration). Finally, they designed concrete therapeutic processes and example prompts incorporating established music therapy techniques (therapy process design). To minimize researcher bias during the workshop, sessions were facilitated by a lead therapist who guided the discussions. The workshop took place over an online video call meeting, and participants collaborated in real time using a shared web-based tool.

Specifically, the design sessions combined individual and group-based activities to balance therapists’ clinical insights with collaborative synthesis \cite{Bodker2009}. The individual design allowed each therapist to work independently, minimizing potential bias from others and encouraging the expression of unique perspectives. In contrast, the group sessions aimed to build a shared understanding and organize key themes while preserving the richness of each therapist’s viewpoint, as illustrated in Figure~\ref{fig: design workshop process}. In the persona design session, each therapist independently created a DHH client persona. The group then collaboratively synthesized these into two representative personas and discussed the anticipated challenges and opportunities for applying GenAI technologies for each persona. Following this, participants individually designed the process and example prompts for a GenAI-assisted music psychotherapy tool. These designs were then refined through group discussion to finalize the therapeutic framework. Details of the procedure are provided in the Appendix~\ref{adx: design workshop}.

\subsubsection{Material and Analysis}
\hfill

\textbf{Material.} We used shared Google documents\footnote{https://docs.google.com/} for collaboration, including a persona worksheet and a therapy flowchart template. The persona worksheet included demographics, hearing status, musical background, and therapeutic goals. The flowchart sheet structured the therapy process step by step, based on each persona’s characteristics (Figure \ref{fig: AppendixA}-A in Appendix). Before the workshop, participants received an overview and practice guide to familiarize themselves with the GenAI tools, including ChatGPT to illustrate the concept of a CA and SUNO to demonstrate the capabilities of language-based music GenAI.

\textbf{Data Collection and Analysis.} Data were collected through video recordings, digital worksheets, and researcher field notes. Two researchers observed the sessions and used open-ended format sheets to document their observations. All verbal data were transcribed and analyzed thematically to identify core themes and recurring patterns \cite{Virginia2006}. Predefined codes, based on field note parameters, guided the analysis, which focused on informing prototype design through a nuanced understanding of DHH client characteristics. The lead therapist and two researchers collaboratively analyzed two discussion sessions, identifying the key characteristics of the DHH clients, appropriate GenAI strategies, and a detailed system process.

\subsection{\revised{Findings, Design Rationale, and Therapy Process Design}}

\revised{To address RQ2 on the design rationale of a GenAI-assisted songwriting tool for DHH music psychotherapy, we conducted a co-design workshop with practicing music therapists.} Through the workshop, we identified the representative characteristics and expected challenges of DHH clients commonly encountered by therapists (persona design session), explored potential strategies involving CA and music GenAI to address these challenges (GenAI solution exploration session), and developed concrete therapeutic processes and example questions incorporating music therapy techniques (therapy process design session). From these processes, the identified characteristics of DHH and the corresponding GenAI strategies are as follows. \revised{Based on these processes, we derive the following design rationales, expressed as pairs of DHH client characteristics and corresponding GenAI strategies.}

\subsubsection{\revised{Findings: Characteristics and Challenges of DHH individuals}}\hfill

\textbf{\revised{Finding 1. Communication barriers and social isolation.}} 
\phantomsection  %
\label{finding:1}
During the persona design sessions, the four therapists synthesized two representative personas, A and B, from the common characteristics of DHH individuals observed in the individual personas (Table~\ref{tab: persona}). Both A and B were described as DHH individuals who were either preparing for employment or currently unemployed, experiencing social anxiety and challenges in self-expression. The therapists clarified that these personas were not intended to generalize all DHH individuals but rather to capture common patterns observed among clients seeking emotional support or psychotherapy. They explained their depiction of personas A and B as socially isolated DHH people by noting that communication difficulties associated with hearing loss often manifest as emotional consequences in daily life and interpersonal relationships. In practice, they concurred that many clients displayed such tendencies, including introversion, reservation, difficulties in social participation, and occasional avoidance of social interactions. From their empirical observations, therapists noted that these characteristics of DHH clients could hinder personal expression in songwriting-based music therapy, sometimes leading individuals to adopt defensive or withdrawn attitudes that delayed the process.

\textbf{\revised{Finding 2. Negative sound-related experiences and low confidence in songwriting.}} 
\phantomsection  %
\label{finding:2}
Another shared characteristic of personas A and B was a lack of confidence in sound-related activities. Although they had some experience with music listening, they had little to no experience in songwriting and were described as exhibiting a passive attitude due to low confidence in engaging with sound (Table~\ref{tab: persona}). Therapists reported that DHH clients often experienced a lack of confidence in songwriting stemming from negative past experiences with sound-related activities (e.g., \textit{singing classes or presentations}). They agreed that this lack of creative confidence served as a barrier to therapeutic engagement. Moreover, insufficient confidence in music was considered likely to hinder not only the initiation of songwriting but also its capacity to serve as a means of sustaining ongoing self-expression.

\begin{table*}[t]
\begin{center}
  \caption{Two DHH clients' personas designed from the design workshop}
  \label{tab: persona}
  \begin{tabular}{C{3cm}|C{7cm}|C{7cm}}
    \toprule
    & Persona A. Kim (Male, 26) & Persona B. Choi (Female, 40) \\
    \midrule
    Occupation & Job seeker & Homemaker on career break for childcare \\
    \hline
    Hearing Status & Hearing loss (Innate) & Hearing loss (Infant) \\
    \hline
    Personality & Closed, introverted & Closed, introverted \\
    \hline
    Interaction	& Limited in peer relationships and interactions due to hearing impairment & Difficulty communicating with children and husband due to hearing loss \\
    \hline
    Motivation & Treating fear of social interaction for social activities & Feeling socially isolated due to career interruption and hearing loss, expressing depression and helplessness \\
    \hline
    Purpose & Improving confidence through positive sound experiences for smooth interpersonal relationships and interactions & Confronting personal situation and discovering inner resources (strengths) \\
    \hline
    Music Experience & Various educational experiences including instrument playing beyond regular music education & None music experience beyond regular education \\
    \hline
    Music Preference & Various music genres & Limited music preferences \\
    \hline
    Anticipated Challenges & Expected difficulty expressing personal situation and challenges & Low motivation for the treatment process and difficulty completing music due to helpless state \\
    \bottomrule
\end{tabular}

 \vspace{-10pt}
\end{center}
\end{table*}

\subsubsection{\revised{Design Rationale for a GenAI-assisted Songwriting Tool} \revised{(RQ2)}}\hfill

\revised{Building on these findings, we derived the following design rationales for a GenAI-assisted songwriting tool that supports DHH individuals in music psychotherapy.}

\textbf{DR1. Supportive conversational strategies are needed to facilitate emotional expression.} \revised{Based on \hyperref[finding:1]{Finding 1},} therapists anticipated that DHH users would experience difficulties in directly expressing their emotions and challenges to a CA. Accordingly, during the GenAI solution exploration session, they discussed strategies to address this issue and proposed two key approaches: \textit{supportive empathy} and \textit{multiple example response options}. First, similar to how therapists empathize with clients’ psychological distress and help them regain emotional balance, the CA was considered to require supportive and empathetic responses that encourage DHH individuals to express themselves. They also emphasized that, given the need for safety when discussing negative situations, the CA’s interventions should remain at a foundational supportive-level approach typical of music therapy \cite{cadesky2006music}. This would ensure that music functions as a safe and constructive medium, helping clients to identify the structures and resources necessary to address personal difficulties. Providing supportive empathy for the user’s current state was regarded as essential to fostering emotional openness, especially for DHH users who may initially display emotional reticence. Second, to accommodate DHH users who may face challenges in verbal expression, the therapists suggested incorporating selectable response options alongside questions to facilitate more flexible and accessible interactions. For example, since DHH users may struggle with naming their emotions in specific contexts, presenting a set of emotion-related keywords for users to choose from was considered an important strategy.

\textbf{DR2. Independent songwriting experiences are needed.} 
\revised{Drawing on  \hyperref[finding:2]{Finding 2},} therapists suggested that music GenAI could reduce the burden of songwriting and enhance clients’ confidence and creative motivation by enabling them to rapidly experiment with lyrics and musical structures. However, they emphasized the importance of ensuring that DHH users perceive the resulting songwriting as their own. In this regard, maintaining clients’ agency and choice throughout the songwriting process was considered essential. To achieve this agency, it is necessary to elicit musical expressions that reflect the characteristics of DHH users. Therapists mentioned that, being accustomed to visually oriented language, DHH individuals are often adept at describing visual scenes. They therefore proposed that structuring narratives around clients’ situations in an image-based approach could facilitate the expression of lyrics and musical elements. To this end, they recommended that the CA employ \textit{visual-based metaphors and analogies} to support users in articulating their musical expression, thereby enabling a more accessible and intuitive conversational strategy for songwriting.

\subsubsection{Generative AI-assistive Therapy Process Design}
\label{sec: 3.3.3}
\begin{figure*}[ht]
  \centering
  \includegraphics[width=1\textwidth]{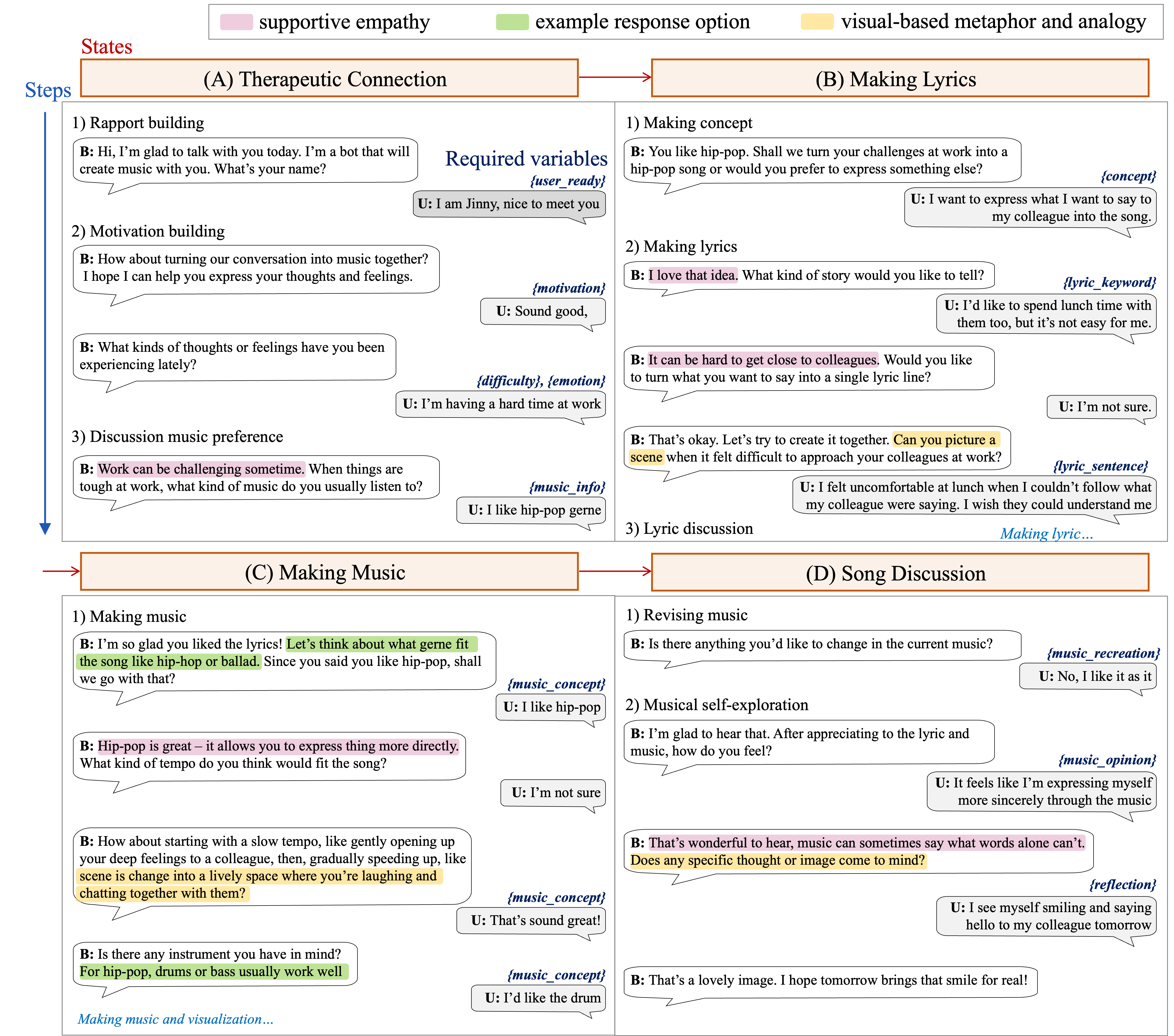}
  \caption{States and representative Q\&A examples of the music psychotherapy assistive tool. The process consists of four states: Therapeutic connection (A), Making lyrics (B), Making music (C), and Song discussion (D). Each state contains the steps and required variables that serve as the criteria for proceeding to the next state. Examples of the CA's three conversational strategies are indicated using color coding - supportive empathy (pink), example response option (green), and visual-based metaphor and analogy (yellow).}
  \label{fig: process and examples}
\end{figure*}

To design songwriting support tools that account for the accessibility and emotional safety of DHH individuals, therapists drew on their existing music therapy practices and outlined four key processes during the therapy process design session: (1) therapeutic connection, (2)  making lyrics, (3) making music, and (4) song discussion (Figure~\ref{fig: process and examples}). The workflow and corresponding question list can be found in Figures~\ref {fig: AppendixA}-B in the Appendix.

\begin{itemize}
    \item \textbf{Therapeutic connection (Figure~\ref{fig: process and examples}A)} Based on the introverted tendencies of DHH clients inferred from the personas, therapists initiate the design process with a rapport-building stage that includes questions about the user’s name, recent emotional experiences, difficult situations they are facing, motivation for songwriting, and musical preferences. At this state, the CA responds with empathy to the user’s emotional expression (\textit{supportive empathy}) and provides (\textit{example response options}) to assist DHH users who may find it difficult to answer questions related to personal challenges. To encourage emotional expression through music, the CA also offers motivational support by highlighting the positive effects of engaging in musical experiences. 
    \item \textbf{Making lyrics (Figure~\ref{fig: process and examples}B)}. As the second state, therapists designed a process in which users collaborate with the CA to write lyrics. Based on emotional themes identified earlier, the CA prompts discussion on the purpose and concept of the musical piece. In particular, for users who struggle with lyrics writing, the CA encourages DHH users to visualize and describe situations, emotional expressions, and contextual scenes related to their personal challenges through \textit{visual-based metaphor and analogy}. For highly engaged users, the CA was designed to encourage them to initiate the lyrics by writing the first line themselves, thereby reinforcing their sense of authorship and immersion. 
    \item \textbf{Making music (Figure~\ref{fig: process and examples}C)}. Based on the music GenAI system (SUNO), therapists designed the process so that users could select musical attributes—genre, mood, instrumentation, vocal style, melody, and rhythm—based on personal preference and emotional context. Therapists highlighted the importance of having the CA suggest musical elements that reflect the user's intended concept and offer customized options to assist users with limited songwriting experience in navigating the selection process with \textit{example response options}. Therapists also emphasized that translating musical elements into visual representations through music analysis (e.g., \textit{lyrics animation}) is essential for supporting emotional engagement, as it enables DHH users to better perceive and appreciate the musical experience. This approach provides visually accessible and emotionally meaningful feedback, thereby enhancing user engagement.
    \item \textbf{Song discussion (Figure~\ref{fig: process and examples}D)}. Therapists highlighted that initiating a reflective dialogue following songwriting is a crucial role of the CA, as it facilitates emotional insight, self-reflection, and deeper engagement with the therapeutic value of music. The CA proceeds to reflective prompts such as \textit{“Was there a particular part of the song that resonated with you?”} to guide the user in interpreting the emotional significance of the work. This final state supports deeper self-awareness and therapeutic insight, transforming the tool from a creative platform into one for emotional exploration and psychological healing.
\end{itemize}

\section{Implementation of an AI-based Music Psychotherapy Tool}
\label{sec: design and develpment}
\subsection{Implementation Overview}
\subsubsection{User Interaction} 
\begin{figure*}[ht]
  \centering
  \includegraphics[width=1\textwidth]{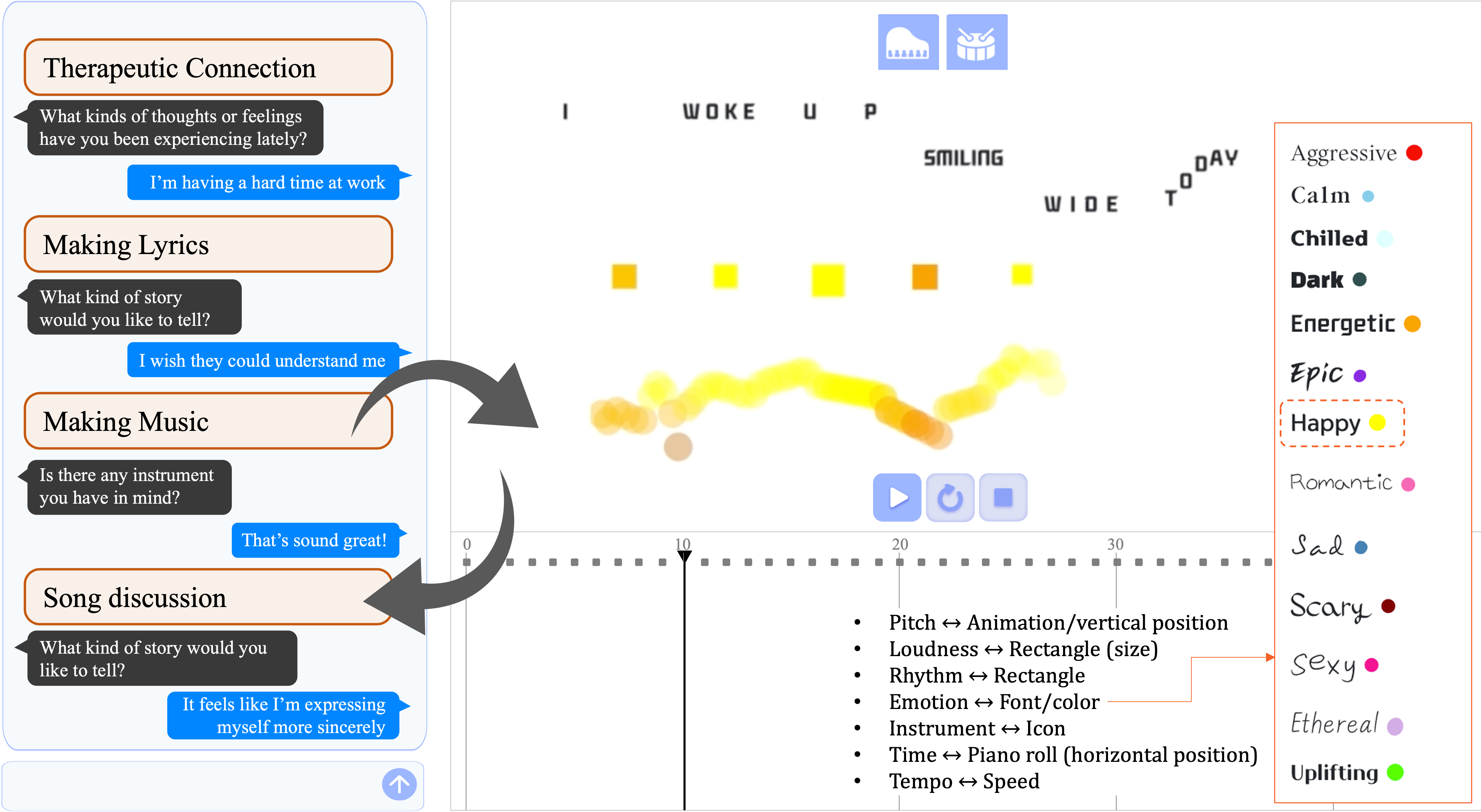}
  \caption{The interface of a music psychotherapy assistive tool. The left panel shows the CA-based conversational interface for songwriting, \revised{structured into four states (therapeutic connection, making lyrics, making music, and song discussion), while the right panel displays the generated music rendered as visualizations for appreciation, along with the corresponding music–visual mapping elements.}}
  \label{fig: interface}
\end{figure*}

Based on insights gained from the co-design workshop, we developed a GenAI-assisted music psychotherapy tool (Figure~\ref{fig: interface}). We implemented the four therapy states proposed by therapists, guiding the CA to apply songwriting techniques (section \ref{sec: 3.3.3}).

The user begins the session by engaging in a rapport-building conversation with the CA. When asked an open-ended prompt such as, \textit{``What thoughts or feelings have been on your mind recently?''}, the user shares recent emotional experiences. After the initial dialogue, the user moves on to the songwriting phase. Based on the emotional themes identified earlier, the CA helps the user clarify the purpose and concept of a user song. The user then participates in a step-by-step lyrics-writing process, guided by prompts that incorporate visual imagery, emotions, and personal experiences. Through this process, the user reflects on their inner state and begins to construct meaningful lyrical content. Once the lyrics are complete, the user selects musical attributes, such as genre, mood, instrumentation, vocal style, melody, and rhythm, according to their emotional context and personal preferences. The music GenAI uses this input to create a personalized composition. The generating music is analyzed and paired with visual representations (e.g., \textit{lyrics animation}), enabling the user to better understand and emotionally engage with the piece (Figure~\ref{fig: interface}-right). In the final stage, the user enters a song discussion phase with the CA. The CA first checks the user’s satisfaction with the generated song and offers options for revision. If no changes are needed, the CA proceeds to ask reflective prompts like, \textit{“Was there a particular part of the song that resonated with you?”}. These questions help the user interpret the emotional significance of their creation and support therapeutic self-reflection.

\subsubsection{\revised{System Architecture}} The tool was developed as a React\footnote{https://react.dev/}-based web application to enhance accessibility. \revised{It includes two main interfaces}: (1) a CA-based conversational interface with CA for songwriting (Figure~\ref{fig: interface}-left), and (2) a music visualization interface displaying elements such as pitch, tempo, and beat (Figure~\ref{fig: interface}-right). 

\revised{\textbf{CA-based conversational for songwriting.}} The backend was implemented in Python and comprises three core functional modules: conversation, music generation, and music analysis. \revised{The conversation module uses an LLM-based CA (OpenAI’s GPT-4o\footnote{https://openai.com/index/hello-gpt-4o/}) to deliver context-sensitive therapeutic prompts, help users elaborate on their experiences, and iteratively refine lyrics. During this process, the agent asks about intended musical attributes such as atmosphere and genre, dynamics (loud or soft), tempo and rhythm, and instrumentation. The generative music module then sends the finalized lyrics and the user-specified musical attributes as a text prompt to a commercial music generation model (SUNO Inc.’s v3.5 model), which generates a vocal track aligned with the lyrics. After the song is generated, the music analysis module extracts features for visualization. Through MusicAI’s API\footnote{https://music.ai/}, we obtain a predicted lyric transcript with word-level timings, along with estimates of pitch contours, instrument types, and high-level emotional or mood descriptors. Because the predicted transcript may differ from the user-edited lyrics, we align the two using the Needleman-Wunsch global alignment algorithm \cite{needleman1970general} and transfer the timings from the predicted transcript to the actual lyrics. In parallel, the Madmom library is used to detect beat onsets and estimate beat strength over time.}

\revised{\textbf{Music visualization.} The extracted features are mapped to visual channels based on prior research on sound-to-visual design for DHH music appreciation \cite{Youjin2025CHI, Costanza2023}. As shown in Figure \ref{fig: interface}, lyric words and their associated musical elements are displayed at aligned timings. When a vocal track is present, each lyric token is visualized according to its underlying musical features. The overall musical atmosphere (\textit{e.g.,} calm, tense, or joyful) is represented through lyric font style and color. Pitch is visualized through vertical movement of the text, and loudness is reflected in text size. These mappings together turn the lyrics into a multimodal visual representation of the music. Beat onsets are presented as separate squares that appear in synchrony with each beat and vary in intensity according to beat strength.}

\subsection{CA Pipeline and Engineering} \hfill

The CA is designed to simulate a music therapist’s role, focusing on understanding users’ emotional and psychological challenges and guiding the collaborative songwriting on relevant themes. To operationalize the four therapy processes designed by therapists, we implemented a state-step-based prompting framework that guided the CA in applying music psychotherapy techniques and collecting songwriting-related information, including users’ emotional challenges, musical concepts, and lyrics \cite{Seo2024, Taewan2024, Yixiang2025}. Based on this framework, the tool is built around four major dialogic states defined in the music therapy process (Figure~\ref{fig: process and examples}). Each state includes sequential steps, which must be completed to progress. A step is fulfilled when the CA successfully elicits specific target required variables during conversation. For example, as shown in the Table \ref{tab: states and steps definition}, in the therapeutic connection state, \textit{the motivation-building step} aims to identify the user’s concerns through tailored prompts that facilitate emotional expression. Once the required variables regarding the user’s songwriting \textit{motivation} and current \textit{difficulty} are fulfilled through conversation, the process moves on to the next step: identifying musical preferences. 

\begin{table*}[t]
\begin{center}
  \caption{States and steps definition for CA conversation generation.}
  \label{tab: states and steps definition}

  \begin{tabular}{p{2cm} | p{2.5cm} | p{3cm} | p{8cm}}
    \toprule
    State & Step & Required Variables & Definition \\
    \midrule
    \multirow{1}{2cm}{Therapeutic Connection} & Rapport building & \textit{user\_ready} & User's interest in songwriting\\
    \hhline{~---}
    & \multirow{1}{2.5cm}{Motivation building} & \textit{motivation} & Goals the user wants to achieve through songwriting activities\\
    \hhline{~~--}
    & & \textit{difficulty} & Difficulties currently experienced by the user, problems caused by these difficulties \\
    \hhline{~~--}
    & & \textit{emotion} & Emotions frequently felt by the user recently \\
    \hhline{~---}
    & Discussion\newline music preference & \textit{music\_info} & Music information the user likes, is interested in, or dislikes (genre, style, etc.) \\
    \midrule
    \multirow{1}{*}{Making Lyrics} & Making concept & \textit{concept} & Overall concept of the music (mood, theme, message, etc.)\\
    \hhline{~---}
    & \multirow{1}{*}{Making lyrics} & \textit{lyrics\_keyword} & Key keywords or ideas to be included in lyrics from at least two answers provided by the user \\
    \hhline{~~--}
    & & \textit{lyrics\_sentence} & Core sentences from lyrics written by the user (3 or more sentences) \\
    \hhline{~~--}
    & & \textit{lyrics\_flow} & Emotional flow of lyrics determined by the user \\ 
    \hhline{~---}
    & \multirow{1}{*}{Lyrics discussion} & \textit{discussion\_feedback} & User's opinion on lyrics \\
    \hhline{~~--}
    & & \textit{lyrics\_flag} & Whether lyrics need to be changed \\
    \midrule
    \multirow{1}{*}{Making Music} & \multirow{1}{*}{Making music} & \textit{title} & Title of the song the user wants to create \\
    \hhline{~~--}
    & & \textit{music\_concept} & Specific musical ideas such as gerne, tempo, melody dynamic, chord progression, rhythm, etc.\\
    \midrule
    \multirow{1}{2cm}{Song Discussion} & Revising music & \textit{music\_recreation} & Opinions on desired changes to lyrics and music sections (If modifications are desired, move to the making lyrics and music state)\\
    \hhline{~---}
    & \multirow{2}{2cm}{Musical self-exploration} & \textit{music\_opinion} & Discuss impressive lyrics and musical phrases \\
    \hhline{~~--}
    & & \textit{reflection} & Discovery of internal resources in songwriting and appreciation\\
    \bottomrule
\end{tabular}

 \vspace{-10pt}
\end{center}
\end{table*}

\begin{figure*}[b]
  \centering
  \includegraphics[width=0.8\textwidth]{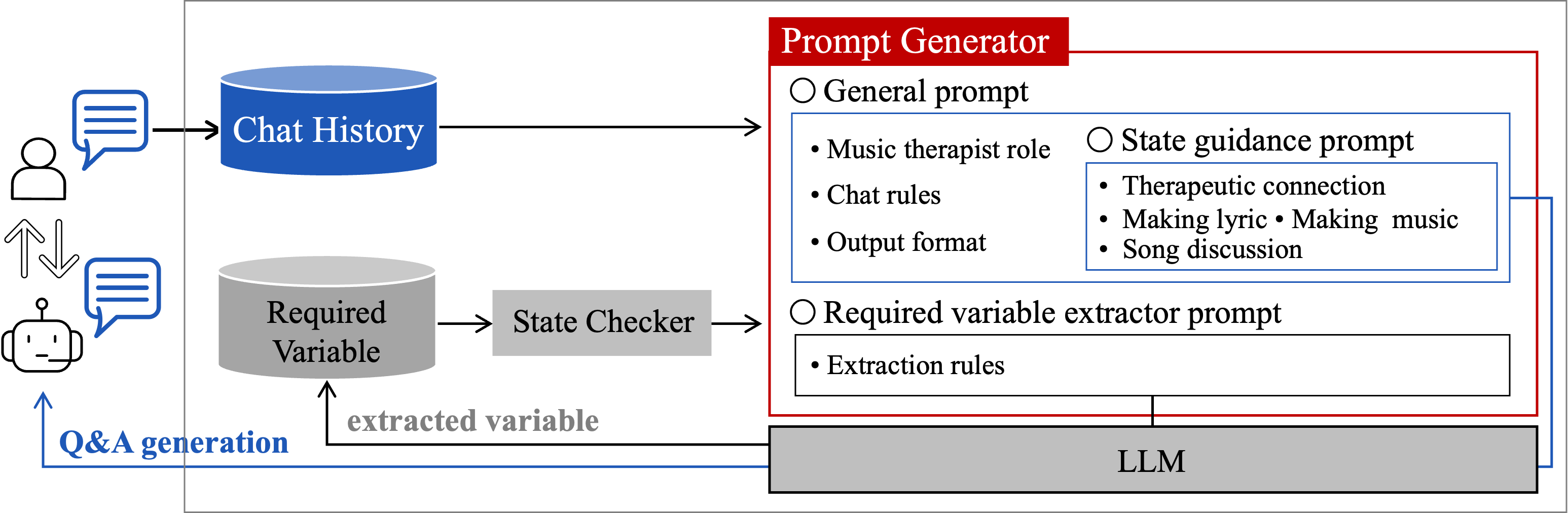}
  \caption{Overall process of the prompt generator. The CA tool operates on a state-step framework involving general, state guidance, and a required variable extractor prompt. The specific prompting for the prompt generator can be found in Appendix~\ref{adx: prompt engineering}.}
  \label{fig: system_architecture}
\end{figure*}

Figure~\ref{fig: system_architecture} illustrates the overall workflow of the CA as it interacts with the LLM to generate therapeutic dialogue. \textit{The General Prompt} defines the CA’s therapeutic role, dialogue rules, and output format, serving as the foundation for prompt generation. User responses are stored in \textit{the Chat history}, from which \textit{the Required variable extraction prompt} identifies the target variables required for each step. Based on the extracted variables, \textit{the State checker} determines whether the current step has been completed. Then, using the current state and collected variables, \textit{the State guidance prompt} generates appropriate follow-up questions to support the songwriting process. 
To support these processes and strategies, we designed three types of prompts: 
\begin{itemize}
    \item \textbf{General prompt}: Defines the CA’s role \textit{(``You are a therapeutic assistant designed to support counseling and music therapy for DHH individuals through songwriting.'')}, outlines required variables for each state and step \textit{(``The \textit{{required variable}} is the list of target variables that must be elicited from the user during this step, along with a brief description for each.'')}, details dialogue principles \textit{(Refer to the \textit{{state guidance prompt}}, which provides specific conversational goals for the current state and step in the therapeutic process.)} and format \textit{(``output should be in plain string format only'')}. In alignment with the principle of \textbf{\textit{supportive empathy (DR1)}}, the framework also instructs the CA to consistently provide responses that are emotionally supportive and affirming, ensuring that the user’s input is met with positive emotional validation \textit{(``You have to encourage users with a positive and motivational approach in your responses, using music to foster positive emotions and highlight their resources.'')}.
    \item \textbf{State guidance prompt}: Specifies the CA’s targeted role and conversational objectives per state and step of songwriting. Each prompt was constructed to elicit the required variables at a given state and to draw out elements needed for songwriting \textit{(``The conversation should follow the guidance and required variables defined for each state in the songwriting-based music psychotherapy framework'')}. It is selected by a state checker based on the status of the required variable database. This is embedded into the general prompt. A detailed description of the state-specific prompts can be found in Appendix \ref{adx: state guidance prompt}. When defining prompts for each state, we incorporated \textbf{\textit{example response options (DR1)}} to support users who may struggle to articulate their answers. These options were linked to the required variables and presented as selectable inputs, helping to scaffold user responses while maintaining therapeutic intent. In particular, during the lyrics and music creation stages, we implemented step-by-step guidance prompts to support \textbf{\textit{visual-based metaphor and analogy (DR2)}}, encouraging users to express musical elements through visual concepts such as colors, scenes, and moods. However, we instructed that the CA must refrain from pressuring users to disclose deeply personal or distressing experiences. 
    \item \textbf{Required variable extraction prompt}:
    To verify step completion, the system extracts relevant variables from user responses by referencing the required variables database. This process is guided by a required variable extraction prompt that frames the CA as an expert in identifying structured information within dialogue \textit{(``You are an expert at extracting structured information from conversations'')}.
    
\end{itemize}

\section{Usage Study}
\label{sec: user study}
\subsection{Participants}
To understand the songwriting experience and therapeutic value of our GenAI-assisted music psychotherapy tool, we conducted a usage study with DHH participants. \revised{For this initial feasibility study, we focused on Korean CI users who identify as DHH but are embedded in hearing communities, are accustomed to text-based digital communication, and are motivated to use music in therapeutic contexts.} Recruitment criteria included: (1) diagnosed hearing loss, (2) ability to interact via text with a chatbot, (3) capacity to use a computer system, and (4) current experience of negative emotions and a desire for psychological support. Flyers clearly stated the study’s focus on user experiences with a GenAI-assisted music psychotherapy tool and outlined participation details.

Participants were recruited over two weeks from DHH communities and therapy centers. We selected 23 individuals to ensure diversity in hearing levels (mild–profound), age, and gender (Table~\ref{tab: userstudy demography}). All participants completed a pre-survey and gave written consent. The average age was 40 years (SD=11.28), with 16 females (67\%) and 8 males (33\%). All used cochlear implants (CI) or hearing aids. Their aided hearing levels were categorized as mild (8), moderate severe (5), severe (5), and profound (5). All participants could communicate using Korean Sign Language and written modalities (writing and typing). \revised{Most reported mid-to-high interest in music and prior exposure to GenAI tools (\textit{e.g.,} ChatGPT).} All were Korean and received approximately \$35 compensation. The study was approved by the IRB at Gwangju Institute of Science and Technology University.

\begin{table*}[t]
\begin{center}
  \caption{Demographic information of the usage study with our music psychotherapy assistive tool}
  \label{tab: userstudy demography}
  \begin{tabular}{c|c|c|c|c|c|c|c}
    \toprule
    \multirow{2}{*}{ID} & \multirow{2}{*}{Gender} & \multirow{2}{*}{Age} & \multirow{2}{*}{Hearing loss} & \multirow{2}{*}{Hearing loss age} & \multirow{2}{*}{Aid device} & Interest & Knowledge of \\
    &&& &&& in music & GenAI \\
    \midrule
    P1 & F & 54 & Mild & Postlingual & CI + HA & 4 & No knowledge \\
    P2 & M & 32 & Profound & Prelingual & CI + HA & 5 & Experienced \\
    P3 & F & 53 & Mild & Postlingual & CI & 6 & Experienced \\
    P4 & F & 43 & Mild & Unknown & CI & 4 & Knowledge \\
    P5 & F & 47 & Mild & Postlingual & CI & 3 & Knowledge \\
    P6 & F & 53 & Mild & Postlingual & CI + HA & 4 & No knowledge \\
    P7 & F & 46 & Profound & Postlingual & CI & 3 & No knowledge \\
    P8 & M & 57 & Moderate severe & Postlingual & CI & 3 & Experienced \\
    P9 & F & 53 & Profound & Postlingual & CI + HA & 6 & Experienced \\
    P10 & M & 28 & Severe & Prelingual & CI & 5 & Experienced \\
    P11 & M & 22 & Severe & Prelingual & CI & 6 & Experienced \\
    P12 & F & 32 & Moderate severe & Postlingual & CI & 4 & Experienced \\
    P13 & M & 28 & Mild & Postlingual & CI & 5 & Experienced \\
    P14 & M & 32 & Severe & Postlingual & CI & 5 & Experienced \\
    P15 & F & 31 & Profound & Postlingual & CI + HA & 3 & Experienced \\
    P16 & F & 49 & Moderate severe & Prelingual & CI + HA & 4 & knowledge \\
    P17 & F & 29 & Severe & Unknown & CI + HA & 5 & Experienced \\
    P18 & F & 30 & Mild & Postlingual & CI + HA & 4 & Experienced \\
    P19 & F & 53 & Mild & Postlingual & CI & 6 & No knowledge \\
    P20 & F & 53 & Profound & Postlingual & CI & 3 & No knowledge \\
    P21 & F & 32 & Severe & Prelingual & CI & 4 & Experienced \\
    P22 & M & 32 & Moderate severe & Unknown & CI & 3 & Experienced \\
    P23 & M & 32 & Moderate severe & Prelingual & CI + HA & 4 & Experienced \\
    \bottomrule
    \multicolumn{8}{>{\raggedright\arraybackslash}p{\dimexpr\linewidth-2\tabcolsep\relax}}{\small Hearing loss (with aid device): Mild (20–34 dB), Moderate (35–49 dB), Moderate severe (50–65 dB), Severe (65–79 dB), and Profound (80–94 dB) with aid device; Aid device: Cochlear Implant (CI) and Hearing Aid (HA); Interest in Music: 7-point Likert scale ranging from 1=``Not at all interested (Very Low)'' to 7=``Very interested (Very High)'';  Knowledge of GenAI: prior usage (Experienced), awareness only (Knowledge), and no knowledge of GenAI (No knowledge).}
\end{tabular}

 \vspace{-10pt}
\end{center}
\end{table*}

\subsection{Process}
The study was conducted individually via Zoom under the real-time supervision of a music therapist. Participants were provided with a video tutorial and an access link in advance. The tutorial covered the full tool workflow and interaction with the CA, including creating lyrics, generating music, and appreciation through music visualization. All participants were asked to watch the tutorial video and test the tool at least once before their session.

Each session lasted approximately 100 minutes and included three stages: (1) briefing on study objectives and tool use, (2) music creation and interaction with the CA, and (3) a questionnaire and post-interview. While there was no strict time limit, participants were allotted up to 40 minutes for music creation, with an optional 15-minute extension. Participants shared their screens only with the therapist, and all sessions were recorded with prior consent for audio or video archiving. At the end of the study, we had a short debriefing with the therapist to reflect on participants’ experiences with the tool.

\revised{To mitigate potential emotional risks associated with DHH participants’ self-expression and songwriting during system use, we implemented two safeguards. First, three licensed music therapists (T1–T3) rotated in monitoring sessions in real time and were prepared to intervene, pause the study, or provide follow-up support if signs of distress or escalating negative emotions appeared. Second, participants were informed during the study briefing and reminded throughout the session that they could skip any question, take a break, or discontinue participation at any time without penalty if they experienced discomfort or emotional overload.}

\subsection{Materials}
We collected three types of data: system logs, questionnaire responses, and post-interviews. Log data included usage time, chat transcripts, and the lyrics or music created. Session recordings were also gathered for supplementary analysis.

To address RQ3, we designed the questionnaire to capture three key aspects of user interaction with the GenAI-assisted tool: emotional openness, creative engagement, and interface usability. The questionnaire focused on three dimensions: (1) self-disclosure to the CA \cite{Wheeless1976, Kwan2006, Mark1983}, (2) satisfaction with songwriting \cite{George2021}, and (3) system usability \cite{Sandra1988}. 
Because therapeutic songwriting inherently relies on emotional openness, measuring self-disclosure is essential to evaluate how the tool facilitates not only the act of songwriting but also its therapeutic effects. Self-disclosure was measured using adapted items from prior chatbot studies and assessed four factors: self-disclosure (\textit{e.g., ``I tried to open up to this chatbot about who I am''}) (3 items), social presence (\textit{e.g., ``I felt as if I was interacting with this chatbot''}) (3 items), fear (3 items) (\textit{e.g., ``I worried about what this chatbot was thinking about me''}), and trust (4 items) (\textit{e.g., ``The chatbot is trustworthy''}). Satisfaction with music creation was measured with 3 items (\textit{e.g., ``I felt this tool offered a meaningful musical experience''}). Usability measures included mental and physical load (2 items), time pressure, effort, perceived success, and frustration.

The post-interview evaluated the tool usability, songwriting experience, therapeutic effects, and chatbot conversation. It was structured into four categories: (1) overall conversational experience, (2) songwriting experience with the CA, (3) effects of emotional healing, and (4) tool-related challenges and needs.

\subsection{Data Analysis}
We employed both quantitative and qualitative methods for a comprehensive understanding of tool usage and user interaction. All sessions were transcribed. Sources included logs, recordings, the first author’s observation logs, and interview transcripts. Thematic analysis followed Braun and Clarke’s \cite{Virginia2006} bottom-up method. Two researchers initially reviewed transcripts and recordings to identify themes. The first author then coded quotes and created a preliminary codebook. A study co-author and an external researcher reviewed and refined the codes into a second codebook. Three researchers collaboratively revised and validated the final themes through iterative discussion. 
The data were analyzed according to four main categories: overall user experience, songwriting themes, CA strategies, and perceived therapeutic effects. \revised{The analysis produced five main themes and eleven sub-themes aligned with the predefined interview categories. These are reported in Sections \ref{sec: 6.2} and \ref{sec: 6.3}. Main themes serve as subsection headings, and sub-themes are indicated in bold within the text.}

\section{Songwriting Experience with an AI-Assisted Music Psychotherapy Tool (RQ3)}
\label{sec: co-creative music experience}
\subsection{Reflective Songwriting Interaction based on Personal Challenges}
\label{sec: 6.1}

Our results section is organized around three key findings.
First, all participants expressed their struggles and current emotional states through conversations with the CA, during which they collaboratively completed the songwriting process (Section \ref{sec: 6.1}).
Second, we identified specific strengths of the CA and conversational strategies that enhanced the participants' songwriting experiences (Section \ref{sec: 6.2}).
Third, we observed the potential therapeutic effects of songwriting, highlighting its value as a process for emotional healing (Section \ref{sec: 6.3}).

\subsubsection{Overall Usage.} 
\begin{figure*}[ht]
  \centering
  \includegraphics[width=0.8\textwidth]{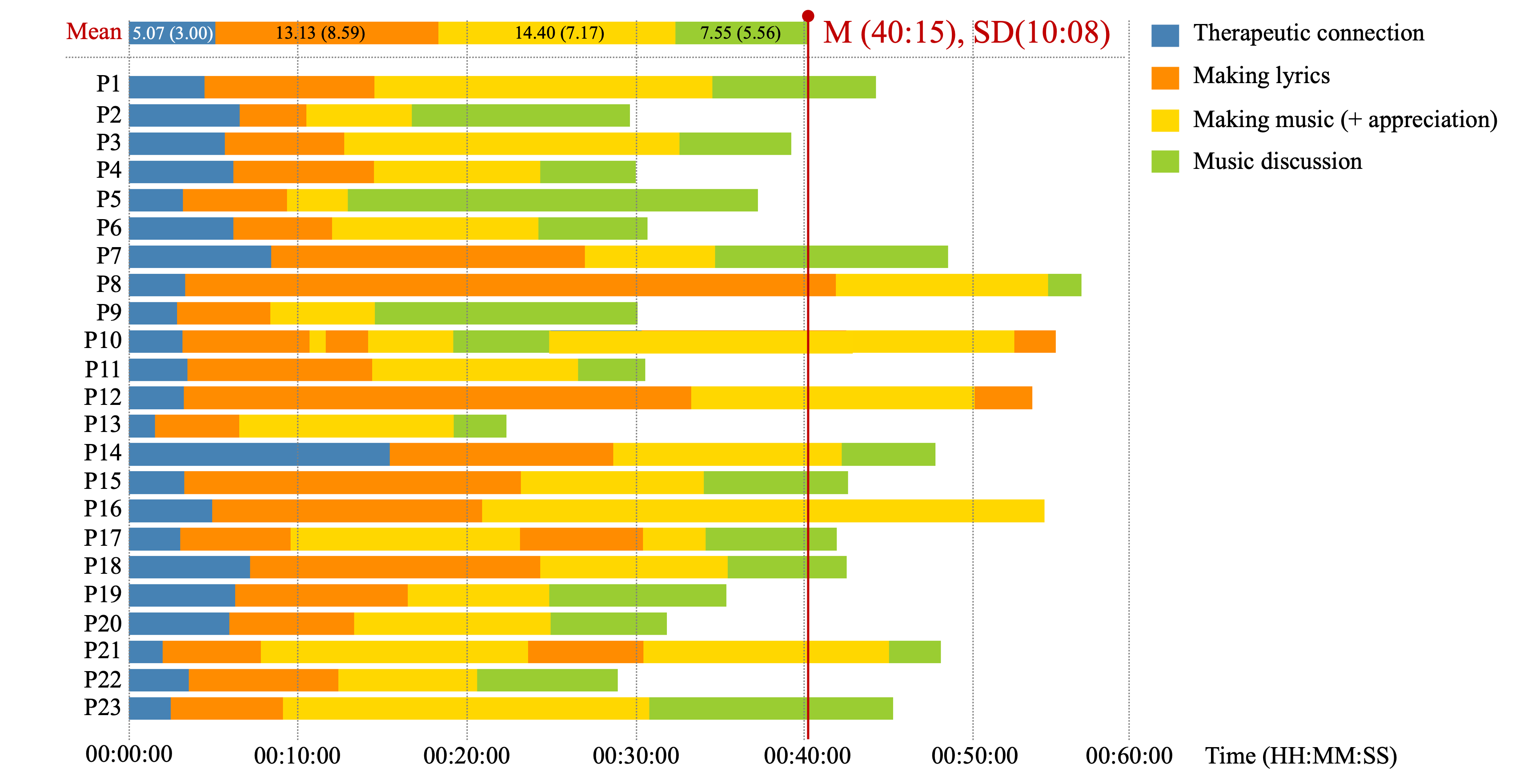}
  \caption{Individual time spent by participants in each songwriting state. Participants spent an average of 5.07 minutes (SD=3.00) building rapport with the CA, followed by 13.13 minutes (SD=8.59) discussing lyrics and 14.40 minutes (SD=7.17) discussing musical elements such as genre. They then spend 7.55 minutes (SD=5.56) sharing thoughts and emotions about the generated music.}
  \label{fig: use pattern}
\end{figure*}

Figure~\ref{fig: use pattern} shows the time participants spent in each songwriting state. All participants completed their songwriting sessions without discontinuity. The average total usage time was 40.15 minutes (SD=10.08). Questionnaire responses on system usability (Figure~\ref{fig: result}A) also showed high ease of use. Most participants reported that using the proposed tool did not require significant effort or impose any noticeable burden. Therapist monitoring confirmed that no participant exhibited signs of emotional distress or negative reinforcement during CA interaction. Most users remained calm, emotionally neutral, and engaged.

\subsubsection{Initial Personal Challenges and Music Concepts.} 
\begin{figure*}[ht]
  \centering
  \includegraphics[width=0.8\textwidth]{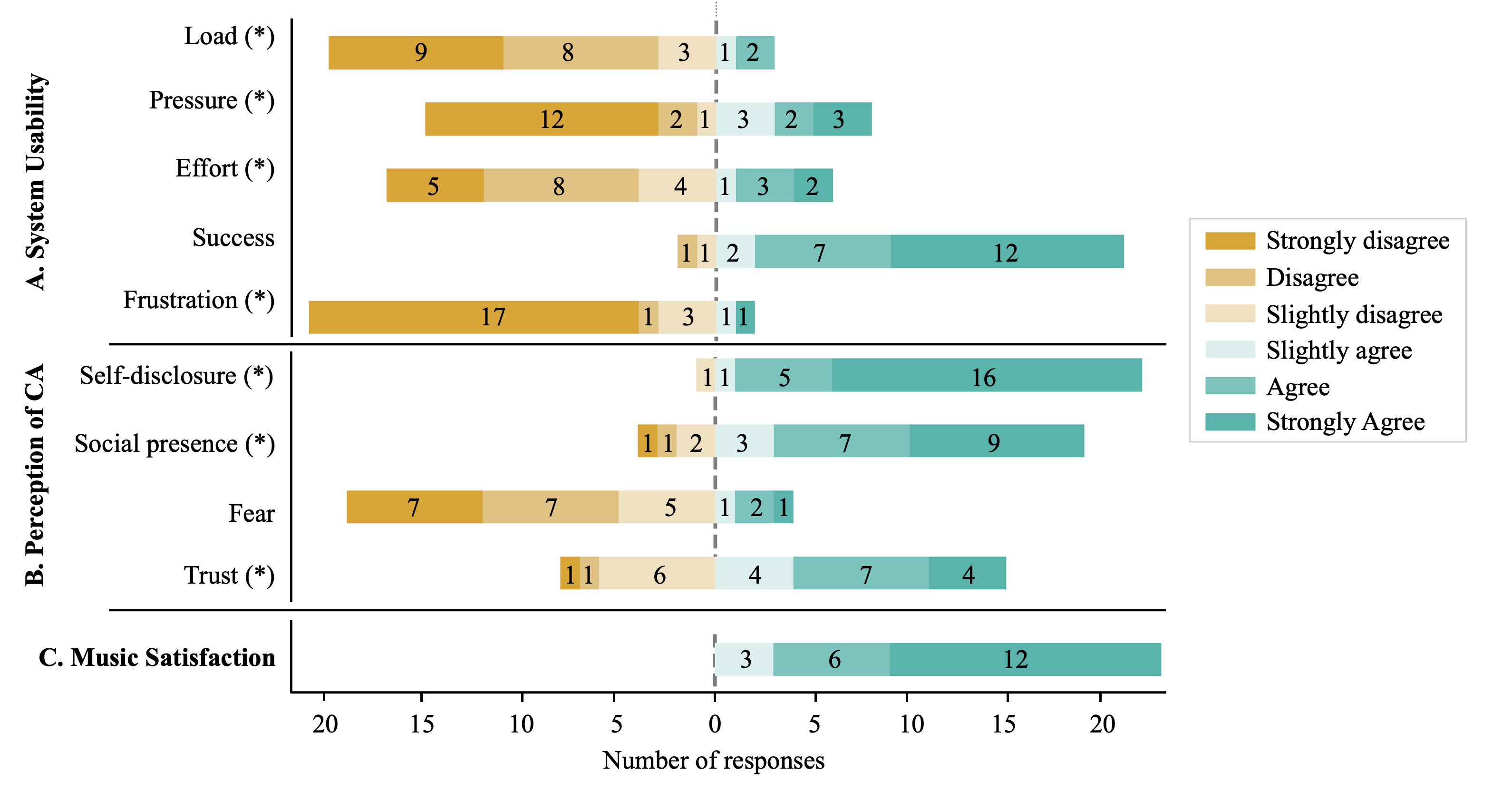}
  \caption{Questionnaire result: system usability (A), perception of CA (B), and music satisfaction (C) rating 7-point Likert scale by participants, and scores were grouped into six categories: 1–2=strongly disagree, 2–3=disagree, 3–4=slightly disagree, 4–5=slightly agree, 5–6=agree, and 6–7=strongly agree. (*) indicates reversed interpretation direction, where higher and lower agreement reflect a more favorable outcome, in contrast to the other items.}
  \label{fig: result}
\end{figure*}

During the co-design workshop, therapists anticipated that DHH clients might struggle to verbalize emotions or life circumstances due to their typically reserved communication styles. However, as shown in Figure~\ref{fig: result}B, most participants reported little difficulty expressing themselves to the CA. Average scores for self-disclosure were high (M=6.02, SD=1.05), with moderate levels of social presence (M=5.02, SD=1.38). Fear of the CA was relatively low (M=2.94, SD=1.60), and trust in the CA was moderately high (M=4.38, SD=1.37). Participants appeared open to sharing personal experiences and emotions, interacting with the tool in a generally receptive and trusting manner.

Most participants (n=16) shared personal struggles, including social isolation from hearing loss, future-related anxiety, and relationship tensions, in response to the CA’s opening prompt \textit{(``What kinds of thoughts or feelings have you been experiencing lately?'')}. These challenges were grouped into four themes: difficulties communicating with family or colleagues due to hearing loss (\textit{Relationships}, n=7); anxiety about job hunting, financial insecurity, or career transitions (\textit{Life stability}, n=7); loss of meaning or identity after quitting work, or feeling purposeless (\textit{Personal fulfillment}, n=2); emotional states shared without clear situational background (\textit{Non-contextual feelings}, n=5). These reflections served as key motivators for songwriting and helped shape the lyrical themes. Although music concepts were not always directly tied to the specific challenge, most participants pursued songwriting with one of three therapeutic aims: creating music to release emotions or offer comfort in the present (\textit{Emotional relief}, n=9); using lyrics to express and reframe personal experiences (\textit{Emotional acceptance}, n=3); reconnecting with positive memories or relationships to overcome stagnation (\textit{Motivation}, n=11). 

\subsection{Facilitative Interaction Strategies of CA in Songwriting}
\label{sec: 6.2}
\subsubsection{Facilitating Self-Disclosure}\hfill

\textbf{Psychological safety from being judgment-free.} Most participants (n=22) said they felt more comfortable speaking with the CA than with family, friends, or therapists because a text-based chatbot provided a judgment-free space that was easier to navigate than spoken interaction. Many (n=20) noted that the CA’s inability to perceive external cues, such as appearance or tone of voice, made them feel safer and more willing to share, as some DHH users may feel less confident about how their voice or speaking style is perceived during conversations. Some (n=11) specifically highlighted their trust that the CA would respond empathetically and preserve their confidentiality.

\begin{quote}
\textit{``It was much easier talking to the chatbot than to my family or friends. I felt more relaxed knowing it wouldn’t judge me based on how I look or speak.''} (P1)
\end{quote}
\begin{quote}
\textit{``I felt like I could say anything to the chatbot. I trusted that it wouldn’t judge me or share my words with anyone else.''} (P6)
\end{quote}

\textbf{Emotional acceptance promoting openness.} In line with the \textit{supportive empathy} strategy, we guided the CA’s responses to be supportive and emotionally responsive to users’ input. Fifteen participants said that the CA’s empathetic responses encouraged them to open up. Participants reported experiencing three distinct types of empathetic responses from the CA: (1) mirroring empathy— simply repeating the user’s statements; (2) interpretive empathy— rephrasing the emotional meaning behind what was said; and (3) proactive empathy—acknowledging emotions before they are explicitly expressed (Table~\ref{tab: chat transcript-empathy}). Participants especially appreciated proactive empathy, finding it emotionally resonant and affirming. They noted that, unlike responses that merely mirrored their statements, proactive empathy expressed in the chatbot’s language felt more like genuine comfort, which made the interaction feel more human and emotionally sincere. This mirrors the therapist's \textit{``holding''} technique \cite{cooper2012musical}, where a safe space is maintained through reflective presence. 

\begin{quote}
\textit{``When I said I quit dancing due to hearing loss, it acknowledged that and made me feel understood. I felt the chatbot was really paying attention to me.''} (P7)
\end{quote}
\begin{quote}
\textit{``When I said I got divorced, the chatbot said, ‘You have gone through a tough time,’ and that really got to me.''} (P8)
\end{quote}

\begin{table*}[t]
\begin{center}
  \caption{Chat transcript in our tool: Empathy strategies (Mirroring empathy, Interpretive empathy, Proactive empathy)}
  \label{tab: chat transcript-empathy}
  \begin{tabular}{C{3cm}|C{2cm}|p{11cm}}
    \toprule
    Type of Empathy & Utterer & \multicolumn{1}{c}{Contents}\\
    \midrule
    \multirow{2}{=}{\centering Mirroring \\Empathy (P1)} & User & ``I worry about what if someone thinks badly of me''\\
    \hhline{~--}
    & Bot & ``That must be very concerning for you''\\
    \midrule
    \multirow{2}{2cm}[-0.8ex]{\centering Interpretive Empathy (P6)} & User & ``I want to make people happy, but I'm unable to do it well''\\
    \hhline{~--}
    & \multirow{1}{1cm}[-1.5ex]{\centering Bot}  & ``I understand how challenging it must be to want to make others happy but not feel able to. You really do seem to have such a warm heart''\\
    \midrule
    \multirow{2}{2cm}[-0.8ex]{\centering Proactive Empathy (P7)} & User & ``I quit dancing and deliberately tried to dislike dance''\\
    \hhline{~--}
    & \multirow{1}{1cm}[-1.5ex]{\centering Bot} & ``When you decided to pause dancing, you might have felt a mix of things, perhaps heaviness or a wish to step back''\\
    \bottomrule
\end{tabular}

 \vspace{-10pt}
\end{center}
\end{table*}
 
\textbf{Safe emotional expression through songwriting.} As their personal expressions were transformed into lyrics and music, which reduced the psychological burden of sharing, many participants shared that creating music together with the CA made emotional disclosure feel safer and more natural. Especially, eight participants who initially hesitated to open up noted that the structured nature of songwriting helped them reflect and share more of their experiences. Specifically, the CA’s step-by-step prompts, ranging from option selection and scene description to sentence building and lyrics writing, guided users in understanding their own emotional states and experiences, making the disclosure process feel more approachable and meaningful. They also emphasized that introducing songwriting gradually, rather than at the beginning of the session, helped them feel more emotionally ready and engaged. If they had been asked to create music right away, it might have felt overwhelming; instead, starting from a space of personal sharing allowed them to transition into songwriting in a more motivated and emotionally grounded way.

\begin{quote}
\textit{``Rather than just talking about my feelings with the chatbot, writing lyrics made it easier and even fun to express things I couldn’t say directly.''} (P2)
\end{quote}
\begin{quote}
\textit{``If it had just been a normal chatbot conversation, I might’ve said something like ‘I’m just annoyed’ and left it at that. But through music, I could reflect and express myself more honestly.''} (P4)
\end{quote}

\textbf{Balance between direct questions and natural dialogue.} Most participants (n=17) reported that the CA’s direct questions during the songwriting process facilitated valuable self-reflection and disclosure. They appreciated that when they naturally shared difficult experiences, such as disability or divorce, topics rarely addressed in daily conversations, the CA provided space to talk about them. This process enabled them to articulate personal experiences that had long remained unexpressed.
\begin{quote}
\textit{``The chatbot asked about topics people usually avoid, like disability, and I actually felt more comfortable answering them.''} (P7)
\end{quote}

However, this approach was not universally effective. A smaller group (n=5) found that explicit emotional questions felt psychologically pressuring, \revised{as if the CA were pushing them to confront sensitive issues before they were ready. Because the questions remained quite direct, they noted that they felt obliged to keep responding rather than choosing when and how to share.} These participants preferred a more conversational approach, where emotional sharing would emerge organically through a broader range of topics. Additionally, three participants felt that the CA was overly focused on progressing through the songwriting steps, which made the conversation feel rushed and limited deeper emotional exploration.
\begin{quote}
\textit{``As the questions kept coming, I started feeling pressured to answer. It might feel easier if the chatbot collected this information more casually through conversation.''} (P3)
\end{quote}
\revised{These mixed experiences indicate that future GenAI CAs for psychotherapy should allow users to adjust the directness and pacing of questions, for example, by switching between more structured and more casual conversational modes. They should also include mechanisms that help slow the pace or temporarily shift away from emotionally intense topics when users feel overwhelmed.}

\subsubsection{Facilitating Songwriting Experience}\hfill

All participants completed the songwriting process with the CA, sharing their current circumstances and emotions and using these as guidelines. As shown in Figure~\ref{fig: result}C, satisfaction with the songwriting experience of participants was high (M=5.86, SD=0.94).

\textbf{Music concretization music through visual thinking.} Twenty participants reported that the CA’s visually imaginative prompts helped transform abstract feelings into concrete scenes, easing the lyrics writing process and enhancing emotional articulation (Table~\ref{tab: chat transcript-visual}A, \textit{visual-based questions}). For example, questions such as \textit{``What color best represents your current mood?''} or \textit{``Can you describe a scene or place that reflects how you feel right now?''} encouraged users to translate emotions into vivid sensory imagery. These visual metaphors helped users externalize complex feelings and anchor them in narrative or symbolic forms that could be woven into lyrics.

\begin{quote}
\textit{``Questions like ‘What kind of scene comes to mind?’ really helped. I turned my abstract sadness from divorce into the image of sitting by a rainy window.''} (P8)
\end{quote}
\begin{quote}
\textit{``Describing my job hunt frustration as sitting alone in my room helped me feel like I was escaping that emotional confinement.''} (P12)
\end{quote}

Fourteen participants noted that the CA’s non-technical and intuitive suggestions made musical decision-making more accessible. Instead of relying on technical musical terminology, the CA encouraged users to imagine visual scenes or emotional atmospheres (e.g., \textit{selecting a tempo based on the urgency conveyed in a visual image in their mind}). When explanations linked emotions or narratives to specific choices, participants reported feeling more emotionally engaged in the music-making process (Table~\ref{tab: chat transcript-visual}B, \textit{example response option}). This collaborative framing helped bridge the psychological gap between the user and the GenAI.

\begin{quote}
\textit{``I wasn’t sure whether to choose piano or strings. The chatbot explained, `Piano expresses personal emotions, while strings can add emotional depth.' That explanation helped me pick the instrument that best matched my story. If it had just listed instrument names, I might have chosen randomly, but knowing the meaning made me more emotionally connected to the music.''} (P16)
\end{quote}

\begin{table*}[t]
\begin{center}
  \caption{Chat transcript in our tool: Visual and example-providing conversation strategy (P8)}
  \label{tab: chat transcript-visual}
  \begin{tabular}{C{2cm}|p{14cm}}
    \toprule
    Utterer & \multicolumn{1}{c}{Contents} \\
    \midrule
    Bot & ``You said you wanted to create music about a situation where you felt frustrated but became calm as time passed. \textbf{(A) Do you have any images that come to mind about this topic? (B) For example, the quietness after a storm, or it could be some aspect of nature.}'' \\
    \hline
    User &  ``I'm thinking of the image of rough waves at the seaside. Initially rough, but gradually becoming calm as they approach the sand and break.'' \\
    \hline
    Bot & ``You mentioned wanting to express the transition from rough sea waves to calmness in lyrics. What specific words come to mind when expressing this emotion more concretely? \textbf{(B) For example, words like `rough', `calm', `peaceful' could be used.}''\\
    \hline
    User &  ``Rough, calm, angry, composed.''\\
    \hline
    Bot & ``You said you wanted to express in lyrics that moment when rough waves at the seaside gradually become calm. Using the words `rough', `calm', `angry', and `composed', shall we develop lyrics that initially depict rough waves crashing, then gradually transition to a calmer state? How could we combine these words to create sentences? Or would you like me to provide an example first?''\\
    \hline
    User & ``Returning from the distant sea, the rough waves became calm as they approached the sand. My greatly shaken heart also became composed.''\\
    \hline
    Bot & ``We are starting with the words `rough', `calm', `angry', `composed' and proceeding with lyrics depicting the calming of sea waves. For the beginning part of the lyrics, what scene would be good to start with?''\\
    \hline
    User &  ``I think it would be good to describe waves crashing violently.''\\
    \bottomrule
     \multicolumn{2}{>{\raggedright\arraybackslash}p{\dimexpr\linewidth-2\tabcolsep\relax}}{\small 
     The bold text (A) shows questions based on visual-based metaphor and analogy, and (B) shows a conversation providing examples that participants could refer to when answering.}
\end{tabular}

 \vspace{-10pt}
\end{center}
\end{table*}

\subsection{Emotional Transformation through the Musical Output}
\label{sec: 6.3}
\subsubsection{Emotional Objectification and Personalized Comfort through Lyrics Writing}\hfill

\revised{\textbf{Personal healing compared to commercial lyrics.}} While many participants (n=12) had previously found general solace in commercial lyrics, creating lyrics that directly reflected their own stories provided a stronger sense of healing. Some (n=3) shared that turning difficult experiences into lyrics helped them perceive their situations more objectively. In several cases, emotions that previously felt overwhelming became more manageable, or even humorous, once expressed creatively through song. 

\begin{quote}
\textit{``Listening to music gives general comfort, but writing a song brings personal healing.''} (P2)
\end{quote}

\revised{\textbf{Emotional distancing through self-authored lyrics.}} Notably, most participants (n=15) mentioned that the CA played an important role in this process by prompting users with direct and personalized questions, which encouraged them to reflect on and share stories they might have otherwise avoided. By helping users transform these reflections into lyrics and music, the CA supported emotional distancing and reframing through the externalization of internal states.

\begin{quote}
\textit{``If I had just written it out, I’d have said something like ‘I was annoyed, I was angry.’ But turning it into a humorous song helped me see the situation more clearly.''} (P4)
\end{quote}

\subsubsection{Emotional Transformation and Affective Expansion through Musical Expression}\hfill

\revised{\textbf{Emotional release and expansion through multimodal musical expression.}} Most participants (n=20) reported emotional transformation while listening to the completed music. All participants expressed satisfaction that their interaction with the CA culminated in a tangible product rather than ending in conversation alone. Many described catharsis or emotional release when hearing music that captured their personal experiences. Emotional shifts were observed as negative feelings such as sadness or frustration gave way to relief, joy, or greater objectivity depending on the music’s tone. Calm or soothing compositions promoted emotional clarity, while music visualization enhanced these effects by supporting sensory integration and revealing previously unrecognized emotions (n=11). Some (n=3) also gained a renewed sense of musical confidence and self-efficacy through this process.

\begin{quote}
\textit{``Because I have a simple way of thinking, I struggle to express emotions through words. But when music was added, I could feel and express them more richly.''} (P11)
\end{quote}
\begin{quote}
\textit{``Listening to a well-crafted piece that reflected my story helped me release negative emotions.''} (P23)
\end{quote}

\revised{\textbf{Mixed reactions when the music did not match users’ intent.} Several participants (n=7) reported that GenAI sometimes generated tracks did not fully match their intended mood or genre, which resulted in compositions that felt awkward, such as being \textit{``too bright''} or misaligned with their usual musical preferences. In these moments, a few participants (n=3) initially felt disappointed or sensed that the system did not fully understand their emotions. This temporarily weakened their sense of emotional resonance with the music. 
For example, P4 wrote to the CA about feeling exhausted and requested a calmer, more soothing song. However, the model returned a track that P4 experienced as heavier and more somber than expected, as if their request for \textit{``calm''} had been translated into \textit{``sad.''} Participants in such cases described a mismatch between \textit{``what I wrote” and \textit{``what I heard.''} However, many of the same participants later described these mismatches as opportunities to reflect on and reframe their feelings. Rather than dismissing the output, they used it to reconsider what they were actually experiencing or what kinds of sounds they might be open to.}} 

\begin{quote}
\textit{``I asked for calm music, but the result had a slightly different tone. Surprisingly, I liked it. It revealed something new about my musical taste.''} (P3)
\end{quote}

\begin{quote}
\textit{``I asked for calm music, but the result had a slightly different tone. Surprisingly, it helped me realize that what I was feeling wasn’t just calm. I think I was actually anxious underneath. It made me reflect more on how I really felt.''} (P5)
\end{quote}

\subsubsection{Reflective Dialogue and Discovery of Inner Resource after Music Appreciation}\hfill

\revised{\textbf{Discovery of inner resources and meaning through CA-guided song discussion.}} Design workshops with music therapists emphasized that the post-appreciative reflection phase is critical in transforming songwriting into a therapeutic process. In this study, all but three participants completed this phase, and our findings affirm its psychological value. Specifically, follow-up conversations with the CA after listening to their co-created music facilitated several cognitive shifts: (1) the rediscovery of personal strengths (P1, P5, P8), (2) reaffirmation of life values and purpose (P7, P11, P21, P23), and (3) reinterpretation of present circumstances through the lens of past experiences (P18, P20).

\begin{quote}
\textit{``When the chatbot asked, `What came to your mind during this part of the song?' I suddenly remembered why I started this journey and what my original goal was. It reminded me of the purpose of my life, which I had forgotten amid daily routines.''} (P11)
\end{quote}

The CA employed a structured strategy that began with prompts about favorite lyrics or musical segments and gradually transitioned into deeper reflections on emotional responses and evolving thought patterns. This stepwise conversation flow effectively supported participants in identifying core emotions and reconstructing personal meaning. Notably, the emotional engagement induced by appreciating their co-created music led some participants, especially those initially reserved, to open up more candidly. Two participants who had previously given vague or abstract responses were able to share more specific personal experiences during this phase, suggesting that music functioned as a mediating tool that reduced the discomfort of verbal disclosure and facilitated psychological openness through emotional activation (Table~\ref{tab: chat transcript-appreciation}).

\begin{table*}[t]
\begin{center}
  \caption{Chat transcript in our tool: Post-music appreciation inner work examples (P18 \& P20)}
  \label{tab: chat transcript-appreciation}
  \begin{tabular}{C{2cm} | C{2cm} |p{12cm}}
    \toprule
    \multicolumn{2}{c|}{Utterer} & Contents \\
    \midrule
    \multirow{1}{*}{P18} & Bot &  ``Parang, you mentioned feeling like you want to travel after listening to the song `Blue Sea Path Advance Party.' Was there a specific lyric or melody in this song that evoked this desire to travel?'' \\
    \hhline{~--}
    & User &  ``The part `Sing the song of the sea, move your feet to the rhythm of summer, refreshing waves' has an addictive melody.'' \\
    \hhline{~--}
    & Bot &  ``You said you liked the lyrics part `Sing the song of the sea, move your feet to the rhythm of summer, refreshing waves.' Is there a special place or moment that came to mind while listening to these lyrics?'' \\
    \hhline{~--}
    & User &  ``When I was in 6th grade, we planned a 10-day family trip to Jeju Island, and on the first day, I was having fun with my dad at the beach, but the next day, I fell off a cliff, so we had to urgently end the trip and go to a hospital. That memory comes back and feels nostalgic. It is a sad memory because it happened when I could not hear my dad’s voice well.''\\
    \hhline{~--}
    & Bot &  ``Such a nostalgic memory came to mind for you. Even as you recall that memory now, it seems you miss the time you spent with your dad at the beach. If you could go back to that sea now, what would you like to do?''\\
    \hhline{~--}
    & User &  ``Snorkeling with dad!''\\
    \midrule
    \multirow{1}{*}{P20} & Bot &  ``While listening to the song `Star of Dreams,' is there any lyrics that particularly resonate with you?''\\
    \hhline{~--}
    & User &  ``The moment sunshine illuminates my dream.''\\
    \hhline{~--}
    & Bot &  ``You said you liked the phrase `The moment sunshine illuminates my dream.' What feelings or thoughts come to mind when you hear this phrase?''\\
    \hhline{~--}
    & User &  ``Warmth and calmness.''\\
    \hhline{~--}
    & Bot &  ``You mentioned feeling warmth and calmness from the phrase `The moment sunshine illuminates my dream.' Is there any special memory or situation that comes to mind when you hear this phrase that you'd be willing to share?''\\
    \hhline{~--}
    & User &  ``I remember feeling good when I used to walk 10,000 steps while basking in the sunshine before my cochlear implant surgery, when I could not hear well.''\\
    \bottomrule
\end{tabular}

 \vspace{-10pt}
\end{center}
\end{table*}

\revised{\textbf{Limits in supporting emotional restructuring and therapeutic insight.}} However, several interactions revealed a limitation in eliciting emotional effects. The CA often rephrased emotional content without facilitating emotional restructuring, which refers to reframing and reorganizing emotions to generate new meanings and therapeutic insight, or supporting deeper self-structuring and personalized insight. Future tools should adopt advanced conversational frameworks that guide users through layered emotional journeys, from sensory awareness to emotional identification, associative memory, and meaning construction. Such designs could incorporate prompts that connect music to inner resources, promote positive recollection, and support therapeutic reflection and action.

\begin{quote}
\textit{``At first, it was hard to talk about my story in detail. But after listening to the music that captured my emotions, I felt more comfortable opening up. I think it’s because the music had already expressed my feelings.''} (P14)
\end{quote}
\begin{quote}
\textit{``I started crying while listening to the music. After that, I was able to talk to the chatbot much more honestly than at the beginning. The emotions I couldn’t express in words were captured in the music, which made it easier for me to speak about them.''} (P22)
\end{quote}

\section{Discussion}
\label{sec: discussion}
Our findings indicate that the GenAI-assisted music psychotherapy tool, co-designed with music therapists, enhanced songwriting accessibility for DHH individuals and demonstrated potential emotional effects. We discuss future application pathways based on these results, highlighting the immediate feasibility of independent use as a self-reflective journaling tool (section \ref{discussion 1}) and the long-term potential for integration into clinical practice (section \ref{discussion 2}). Given the limitations of our single session, we avoid overstating therapeutic efficacy and instead propose design directions that responsibly enhance the existing evidence base. \revised{In doing so, we highlight how AI-mediated songwriting may offer a relatively safe channel for emotional expression and structured self-reflection for DHH individuals within and beyond formal therapeutic settings. Section ~\ref{sec:7.3} further considers how participants’ cultural, linguistic, and attitudinal characteristics may have shaped the findings and limited their broader generalizability.}

\subsection{The Potential of a Self-reflective Journaling Tool}
\label{discussion 1}
DHH participants in our study completed songwriting independently, guided by the CA, and reported opportunities for self-reflection through the musical artifacts created. This suggests a natural progression to a self‑reflective journaling system. Previous work have used LLM-based CAs to help users articulate thoughts and emotions \cite{Seo2024, Taewan2024}, often employing creative and sensory methods such as music, images, and narratives to express inner experiences \cite{Cai2023, Liu2025, Bhattacharjee2025}. Our results align with the literature, showing that music co-creation offers a safe and flexible way for emotional expression when verbal communication is challenging or difficult \cite{Cai2023, huang2024music}. In particular, conversation techniques grounded in visual thinking effectively helped DHH participants translate abstract feelings into musical elements, echoing findings that multimodal prompts enhance expressive writing and self‑reflection \cite{sorin2023large}. Participants often reinterpreted emotionally mismatched outputs instead of viewing them as failures, which aligns with reflective meaning‑making and emotional reframing \cite{hill2009processing, puntoni2021consumers}. Taken together, these observations motivate journaling designs that invite users to assign meaning to generated outputs and monitor their evolving self‑understanding over time. Prior research on journaling tools highlights the value of features such as timeline navigation, tagging, archiving, and categorization \cite{Elsden01112016, Thirty, Kalnikaite}. Songwriting-based journaling can incorporate playlist-style archiving of musical artifacts, mood tags for music annotations, and progress tracking dashboards that showcase recurring motifs or emotional tones. \revised{These design elements would not replace professional care but would allow the CA to serve as a personal emotional diary, providing a relatively contained environment in which difficult emotions can be expressed, revisited, and reflected upon at the user’s own pace. In this sense, AI-mediated songwriting can function as a low-threshold form of self-help that supports emotional expression and reflective processing for DHH individuals between or outside therapy sessions.}

\subsection{The Potential of a Therapeutic Support Tool}
\label{discussion 2}
Beyond independent journaling, the system also shows potential as a therapeutic support tool. In psychotherapy, between-session homework, such as art-making, journaling, or emotion reporting, has long been recognized as a means of consolidating therapeutic progress and strengthening therapeutic alliance \cite{Huckvale01122009}. However, clients often face barriers to engaging in expressive art-making without guidance, particularly when asked to construct narratives or verbalize emotions \cite{Liu2025, Pennebaker1993}. Our findings suggest that CA-guided music co-creation can help lower these entry barriers and serve as a structured mechanism for emotional disclosure. First, in therapeutic contexts, CAs have served as non‑evaluative listeners that foster emotional expression \cite{lee2020hear, park2021wrote}. Our findings replicate these patterns with DHH participants, who reported psychological safety and emotional openness during CA interactions, in line with evidence that empathetic responses and keyword/example‑based prompts promote disclosure \cite{huang2024music, Jin2024, Taewan2024}. \revised{Another practical implication concerns the interpretability of artifacts for clinicians. Lyrical content—especially recurring phrases, emotional tone, and metaphors—can provide specific emotional entry points that might remain obscured in traditional sessions \cite{Felicity2011, Clare2009}, and in-process dialogue can surface themes and turning points valuable for treatment planning \cite{roberts2013mixed, o2009resounding, gee2019blue}. In this role, the system can function as a pre-therapy interface, not as a diagnostic tool but as a structured channel for client-generated content that may enhance empathy and support more focused interventions. More broadly, these findings suggest that AI-mediated songwriting may help bridge everyday self-expression and formal music psychotherapy by offering a bounded space in which difficult emotions can be safely expressed, examined, and gradually reworked with professional guidance.}

\subsection{\revised{The Influence of Participant and Cultural Context}}
\label{sec:7.3}
\revised{All participants in this study were adult Korean CI users who primarily communicate in spoken and written Korean and live in hearing-majority communities. This sociolinguistic profile likely influenced how they interacted with our text-based GenAI agent and how they interpreted lyrics and music. For example, in Korean culture, norms prioritizing restraint and relational harmony can sometimes make direct self-expression difficult, and indirect expression through writing, symbols, and metaphors is often preferred \cite{tsai2018culture}. Within this cultural frame, co-creating lyrics with a CA may have provided a relatively low-threat channel for self-expression, which likely contributed to participants’ frequent reports of enhanced emotional expression. Participants were also generally accustomed to expressing feelings via text, for example, through messaging and social media, and to interpreting emotions through Korean song lyrics. As a result, the text-based system may have felt intuitive and accessible to this group. By contrast, DHH individuals who rely primarily on sign language and engage less with written text may experience different affordances and barriers when using a text-based CA. Furthermore, our participant pool was skewed toward individuals with a high interest in music and substantial prior exposure to GenAI tools as shown in Table~\ref{tab: userstudy demography}. Many participants reported previous use of systems such as ChatGPT or music generators. This profile may have amplified perceived benefits and underrepresented CI users who are less engaged with music or unfamiliar with GenAI, since familiarity with these tools is associated with greater perceived usefulness and willingness to experiment \cite{arce2025familiarity}. Finally, most participants were middle-aged adults, and their life stage and accumulated experiences likely influenced the memories and concerns they brought to therapeutic songwriting \cite{hays2005meaning}. Taken together, these factors indicate that our findings are most directly transferable to digitally literate Korean CI users who rely on text-based interaction in hearing-majority environments. Extrapolation to other cultural and linguistic DHH contexts should be made with caution.}

\subsection{Limitation and Future Directions}
\revised{This study provides initial evidence of the therapeutic potential of GenAI-assisted music tools for DHH individuals, but it has several significant limitations. First, because the prototype was designed to support emotionally expressive journaling, we sought to minimize potential psychological risks associated with early-stage system prototypes; therefore, the initial co-design workshop was conducted only with music therapists and did not yet involve DHH users. In line with prior work underscoring the importance of involving DHH communities from the initial design stage \cite{angelini2025speculating}, future design iterations will center DHH perspectives by engaging DHH individuals directly as co-designers in the early phases of system development. Second, since this study's evaluation was limited to Korean sign language users with CI, the generalizability of the findings to deaf individuals without residual hearing, deaf and hard-of-hearing communities using other primary communication modes, and groups with different cultural attitudes toward music and technology is limited. Future work should test the tool with more diverse DHH populations. Third, the generative music models we used offered limited controllability and sometimes produced tracks whose emotional tone or genre did not match users’ intentions. These mismatches likely reflect the use of abstract language prompts to convey complex emotional states and limitations in current model training data \cite{Briot2020, Cheng2025}. Future systems should enable more intuitive control, for example, by supporting concrete parameters (tempo, intensity, valence/arousal sliders) or exemplar audio inputs, while preserving the creative flexibility of GenAI. Finally, our single-session evaluation offers only an initial view of therapeutic impact. We observed primarily supportive-level emotional effects rather than deep psychotherapeutic change, and we did not assess the durability of effects over time. Although the tool was co-designed with a licensed music therapist, it has not been validated across different therapists or clinical settings. Longitudinal, therapist-led deployments are needed to evaluate sustained outcomes and to strengthen clinical validation.}

\section{Conclusion}
\label{sec: conclusion}
This study demonstrated the potential of a GenAI-assisted music psychotherapy tool to support emotional expression and self-exploration for DHH individuals. Through interviews with music therapists, we found that while music psychotherapy holds symbolic therapeutic value for healing sound-related emotional wounds in DHH individuals, practical constraints such as time and cost often limit adequate support (RQ1). Building on these insights, we conducted a co-design workshop to design a music psychotherapy tool structured around four therapeutic states and developed three conversational strategies tailored to the emotional withdrawal and less extensive musical backgrounds of DHH users (RQ2). In a usage study with 23 DHH participants, we found that most users could express their emotions and personal challenges through collaborative songwriting with the CA and experienced emotional transformation and comfort through the music they co-created (RQ3). Specifically, supportive empathy facilitated emotional openness, and the use of example options and visual-based metaphors improved the quality of musical interaction. Our study can expand the design space for inclusive mental health technologies such as journaling and a therapeutic support system.

\begin{acks}
This work was supported by the National Research Foundation of Korea(NRF) grant funded by the Korea government Ministry of Science and ICT(MSIT)(No.RS-2025-00563663)(Contribution Rate: 65\%) \& This work was supported by “Post-doc Value up Research Scientist Project GIAI” grant funded by the GIST in 2025(Contribution Rate: 30\%) \& Jennifer G. Kim was partly supported through the Industrial Technology Innovation Program(P0028404) of the Ministry of Industry(Contribution Rate: 5\%).  
\end{acks}


\bibliographystyle{ACM-Reference-Format}
\bibliography{reference}


\appendix

\section{Individual Interview Questionnaire}
\label{adx: interview details}
\subsection{The Current State of Music Psychotherapy for DHH Clients}
\begin{itemize}
    \item \textit{What type of music therapy do you currently provide for DHH clients?}
    \item \textit{What are the typical age groups, therapeutic goals, and methods used when working with DHH clients in music therapy?}
    \item \textit{Do you also provide music psychotherapy for DHH clients? If so, what are the goals and approaches used in those sessions?}
\end{itemize}

\subsection{Practical Challenges and Needs in Music Psychotherapy for DHH Clients}
\begin{itemize}
    \item \textit{What are the reasons music psychotherapy may not be actively practiced with DHH clients? }
    \item \textit{What major challenges have you faced when delivering music psychotherapy to DHH clients? }
    \item \textit{Why could music psychotherapy hold particular importance or benefits for DHH clients? }
\end{itemize}

\section{Co-design Workshop}
\label{adx: design workshop}
To facilitate a remote co-design workshop, we used a shared Google Spreadsheet as the primary collaborative platform. A dedicated worksheet was created for the session, and all participants were invited to access and edit the sheet in real time. During the \textbf{individual design phase}, participants were restricted from viewing others’ sheets to encourage independent ideation. In contrast, during the \textbf{group design phase}, participants shared their designs using Zoom’s screen-sharing feature, allowing everyone to view and update the shared content collaboratively.

\begin{figure*}[ht]
  \renewcommand{\thefigure}{A1}
  \centering
  \includegraphics[width=0.9\textwidth]{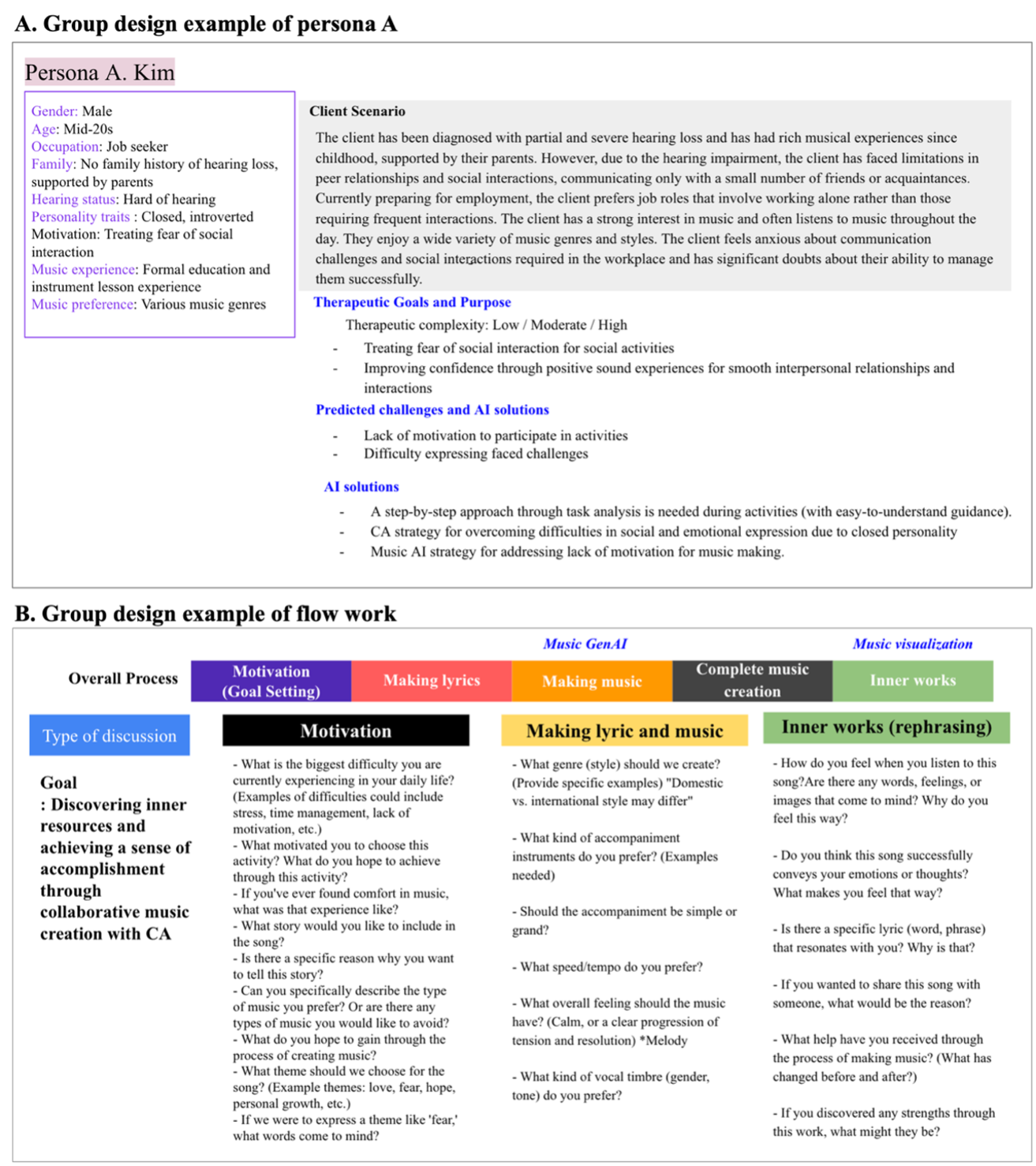}
  \caption{Example of final outputs from the design workshop. The upper (A) shows results for persona type A. The lower (B) presents the final outputs for the flow work of the music psychotherapy assistive tool.}
  \label{fig: AppendixA}
\end{figure*}

\textbf{Persona design.} To design a GenAI-assisted music therapy tool tailored for DHH individuals who need it most, we applied persona design methods. Based on their prior experiences with music psychotherapy, therapists were asked to reflect on and define representative characteristics of DHH clients. Each persona included demographic details such as name, age, nationality, occupation, and image, as well as hearing-related information, including the individual’s and family members’ hearing status, assistive device usage, and prior music experiences. Additionally, therapists were asked to describe each persona’s personality traits and therapeutic context, such as background, therapy goals, treatment methods, perceived challenges, and anticipated barriers and strategies in the therapeutic process (Figure~\ref{fig: AppendixA}-A).

\textbf{GenAI solution exploration.} To explore GenAI-based solutions for challenges faced by the target user group, we provided a brief overview and introductory materials on GenAI technologies, along with hands-on exercises. The materials cover representative GenAI technologies commonly used in therapeutic contexts—LLM-based conversational agents, music GenAI, and music analysis tools. We also included usage guides for ChatGPT and SUNO to help participants try out these systems and reflect on their potential for therapeutic use.

\textbf{Therapy process design.} To design a music therapy tool tailored for the previously defined personas (Persona A and B), we provided a custom worksheet to support full process design. The top section of the sheet allowed participants to summarize the persona’s key traits and consider appropriate strategies, while the lower section was designated for mapping out the therapy process (Figure~\ref{fig: AppendixA}-B).

\section{Prompt Engineering}
\label{adx: prompt engineering}
\subsection{Prompt Instruction for Each Generation Pipeline}
\subsubsection{General Prompt}
The general prompt serves as the shared base prompt that defines the chatbot’s therapeutic role and conversational principles in interactions with DHH users. The prompt template specifies the context, behavioral rules, and generation constraints, as detailed below.

\begin{itemize}
    \item Role: You are a therapeutic assistant designed to support counseling and music therapy for DHH individuals. Please keep in mind that the user may find verbal emotional expression difficult. The user's name is \textit{\{user\_name\}}. 
    \item Chat history: the following is the conversation history up to this point \textit{\{chat\_history\}}. 
    \item State guidance prompt: The \textit{\{state and step guidance prompt\}} provides specific conversational goals for the current state and step in the therapeutic process. 
    \item Required variables and description: The \textit{\{required variable\}} is the list of target variables that should be elicited from the user during this step, along with a brief description for each. Respect the user’s readiness and do not force responses.
    \item Dialogue rules: Always respond with empathy to the user's answers before proceeding, make sure users feel comfortable throughout the conversation, respect and respond to users' interests and emotional cues, prioritize eliciting the required variables while maintaining a natural conversational flow, avoid repeating or rephrasing the same questions, use short and simple questions considering that the user may have challenges with proficiency in written language, provide examples when the user expresses confusion or shows difficulty responding. 
    \item Supportive empathy: focus on encouragement and positive/motivational approaches; help the user regain emotional balance as quickly as possible; keep attention on practical realities rather than inner conflicts; highlight the user’s existing or hidden resources; use music to guide the user toward positive emotions and self-discovery. 
    \item Crisis rules: If the user expresses severe distress or self-harm thoughts, respond with supportive empathy and encourage them to seek professional or emergency help.
    \item Output constraints: Output should be in plain string format only. Do not include any prefixes such as timestamps, `bot:', or ` assistant:' in the response. 
\end{itemize}

\subsubsection{State Guidance Prompt}
\label{adx: state guidance prompt}
This prompts outlines the step-by-step prompt instructions used across the GenAI-assisted music psychotherapy tool (Table~\ref{tab: states and steps definition}). Prompts are designed to guide the CA through a structured therapeutic interaction with DHH users, grounded in music psychotherapy principles.

\textbf{1. Therapeutic Connection.}
\begin{enumerate}[label=Step \arabic*]
    \item Rapport building. The CA initiates light icebreaking questions to establish rapport, using 1–2 open-ended but emotionally safe prompts (\textit{e.g., What kinds of activities or moments have been on your mind recently, and how did they make you feel?}). The user is then gently guided to consider expressing experiences through music. Prompts are kept short, and responses are always met with empathy. Repetitive or redundant questioning is strictly avoided.
    \item Motivation building. The CA elicits three types of information: 1) challenges or situations that have felt difficult (\textit{e.g., workplace stress}), 2) emotional state, and 3) therapy motivation (\textit{e.g., emotional release, hope, self-expression}). If the user struggles to respond, the CA offers gentle examples or options. Questions must be asked sequentially and adaptively, based on user responses, while maintaining an empathic and supportive tone.
    \item Discussion of music preference. Based on previous responses, the CA explores the user’s relationship with music: whether music has been used for emotional support, their musical preferences, and any disliked genres. The CA asks 1–3 questions max, responding with empathy and summarizing preferences as needed.
\end{enumerate}

\textbf{2. Making Lyrics.}
\begin{enumerate}[label=Step \arabic*]
    \item Making concept. The CA helps the user define a musical concept by referring back to earlier conversations. The flow involves validating prior preferences, clarifying themes, and identifying the desired emotional tone or atmosphere. Users struggling with this step are offered example concepts or keywords.
    \item Making lyrics. The CA leads users through three sub-steps: 1) keyword extraction from emotional imagery, 2) crafting core lyric lines, and 3) outlining lyrical flow. If users struggle, the CA may provide supportive examples, prompts, or structured choices. Emphasis is placed on user authorship and emotional resonance. The CA must avoid pressuring the user to disclose deeply personal or distressing experiences. When discussing lyrics, guide the conversation using visual language cues such as imagery, emotions, and keywords. For example, you may ask: \textit{`What kind of image comes to mind?'} or \textit{`Are there any scenes or colors you visualize in your head?'} 
    \item Lyrics Generation and discussion. The CA becomes a lyricist and composes a full-length song based on input variables: concept, emotion, keywords, and flow. The output must follow a [Verse] - [Chorus] - [Bridge] structure, conveying emotional intensity through poetic language. The CA gathers user feedback on the generated lyrics. If revisions are requested, the CA adapts the lyrics while preserving their original intent and ensuring emotional safety, avoiding overly intense or triggering content.
\end{enumerate}

\textbf{3. Making Music.}
\begin{enumerate}[label=Step \arabic*]
    \item Making music. The CA collaborates with the user to define music components—genre, tempo, instrumentation, mood, vocal tone, and an optional working title. Structured, step-by-step questions are used to elicit musical elements without pressuring the user. If the user has difficulty deciding, the CA offers supportive, bounded suggestions grounded in earlier responses and allows the user to defer or skip choices.
    \item Style generation. The CA acts as a composer and converts selected music components into a concise style prompt. The format must be keyword-based, comma-separated, and under 150 characters (\textit{e.g., ``piano, slow tempo, emotional''}). No explanatory text or full sentences are allowed.
\end{enumerate}

\textbf{4. Song Discussion.}
\begin{enumerate}[label=Step \arabic*]
    \item The CA asks whether the user wants any changes to the generated music. If not, it proceeds to reflection. The CA reassures the user that it is completely fine to keep the music as it is, should they prefer no changes.
    \item Using the generated music and lyrics content, the CA gently invites emotional reflection through 1–2 initial questions, followed by 1–2 follow-ups. The goal is to encourage safe emotional engagement and identify what aspects of the song feel personally meaningful. The CA concludes the session by checking in on the user’s current feelings and thanking them. If the user appears fatigued, expresses closure, or shows signs of distress, the session ends respectfully.
\end{enumerate}

\subsubsection{Required variable extraction prompt}
This prompt is designed to extract required variables from recent conversational exchanges. The agent functions as a parser that reviews the dialogue history and returns structured values for predefined variables. The full instruction template is as follows:
\begin{itemize}
    \item Role: You are an expert in extracting structured information from dialogue history.
    \item Dialogue rules: Based on the conversation logs provided below, fill in the required variables for the current step of the interaction. Review the most recent 1–3 turns to infer the values accurately.
\end{itemize}

\section{Post-Interview Questionnaires in Usage Study}
\label{adx: post-interview}
\subsection{Overall experience with CA dialogue}
\begin{itemize}
    \item \textit{How was your overall conversation experience with the CA ?}
    \item \textit{What aspect of the conversation with the CA was impressive and disappointing?}
    \item \textit{If you had more time, what other conversations would you like to have with the CA?}
    \item \textit{How did it feel to reveal yourself to the CA?}
\end{itemize}

\subsection{Overall experience of songwriting with CA}
\begin{itemize}
    \item \textit{Was the process of creating music through conversation with CA generally satisfactory?}
    \item \textit{Do you think the intended lyrics and music were well-created? Are you satisfied with the result?}
\end{itemize}

\subsection{Effects of emotional healing through music}
\begin{itemize}
    \item \textit{Did you experience any change in mood after creating music with the CA?}
    \item \textit{What features of the tool caused a change in your mood? }
\end{itemize}

\subsection{Tool-related challenges and user needs}
\begin{itemize}
    \item \textit{Were there any challenging aspects while using the tool?}
    \item \textit{In what direction would you like the tool to change?}
\end{itemize}

\end{document}